\documentclass[apj]{emulateapj}\usepackage{apjfonts}

\shorttitle{Infrared SBF Distances Using WFC3}
\shortauthors{Jensen et al.}
\newcommand{\kmsmpc}{km\,s$^{-1}$\,Mpc$^{-1}$}
\newcommand{\ho}{$H_{\rm 0}$}
\newcommand{\hst}{\textit{HST}}
\newcommand{\Mbar}{$\overline M$}

\newcommand{\Mbarj}{$\overline M_{110}$}
\newcommand{\mbarh}{$\overline m_{160}$}
\newcommand{\Mbarh}{$\overline M_{160}$}
\newcommand{\mM}{$(m{-}M)$}
\newcommand{\colorbar}{$(\overline m_{110}\,{-}\,\overline m_{160})$}
\newcommand\zacs{\ifmmode z_{850}\else$z_{850}$\fi}
\newcommand\gacs{\ifmmode g_{475}\else$g_{475}$\fi}
\newcommand\gminz{\ensuremath{(g{-}z)}}
\newcommand\jminh{\ensuremath{(J{-}H)}}
\newcommand\gz{\ensuremath{(g_{475}{-}z_{850})}}
\newcommand\JH{\ensuremath{(J_{110}{-}H_{160})}}
\newcommand\gzacs{\gz}
\newcommand\Hir{\ensuremath{H_{160}}}
\newcommand\Jir{\ensuremath{J_{110}}}

\begin{document}

\title{Measuring Infrared Surface Brightness Fluctuation Distances \\ 
with \emph{HST} WFC3: Calibration and Advice\altaffilmark{1}}
\altaffiltext{1}{Based on observations with the NASA/ESA \textit{Hubble Space  Telescope}, 
obtained at the Space Telescope Science Institute, which is operated by the 
Association of Universities for Research in Astronomy, Inc., under NASA contract 
NAS 5-26555. These observations are associated with Program \#11712, \#11570, and \#11691.}

\author{Joseph B. Jensen}
\affil{Utah Valley University, Orem, Utah 84058, USA; {\tt jjensen@uvu.edu}}
\author{John P. Blakeslee}
\affil{NRC Herzberg Astrophysics, Victoria, British Columbia, Canada}
\author{Zachary Gibson}
\affil{Utah Valley University, Orem, Utah, USA}
\author{Hyun-chul Lee}
\affil{The University of Texas Rio Grande Valley, Edinburg, Texas, USA}
\author{Michele Cantiello and Gabriella Raimondo}
\affil{INAF--Osservatorio Astronomico di Teramo, Teramo, Italy}
\author{Nathan Boyer}
\affil{Brigham Young University, Provo, Utah, USA}
\author{Hyejeon Cho}
\affil{Department of Astronomy and Center for Galaxy Evolution Research,
Yonsei University, Seoul 120-749, Korea}

\begin{abstract}
We present new calibrations of the near-infrared surface brightness fluctuation
(SBF) distance method for the F110W (\Jir) and F160W (\Hir) bandpasses of the
Wide Field Camera~3 Infrared Channel (WFC3/IR) on the \emph{Hubble Space Telescope}. 
The calibrations are based on data for 16 early-type galaxies in the Virgo
and Fornax clusters observed with WFC3/IR and are provided as
functions of both the optical \gz\ and near-infrared \JH\ colors.
The scatter about the linear calibration relations for the luminous red galaxies
in the sample is approximately 0.10 mag, corresponding to a statistical error 
of 5\% in distance. Our results imply that the distance to any suitably bright 
elliptical galaxy can be measured with this precision out to about 80~Mpc in a 
single-orbit observation with WFC3/IR, making this a remarkably powerful instrument 
for extragalactic distances. The calibration sample also includes much bluer and 
lower-luminosity galaxies than previously used for IR SBF studies, revealing 
interesting population differences that cause the calibration scatter to increase 
for dwarf galaxies. Comparisons with single-burst population models show that, 
as expected, the redder early-type galaxies contain old, metal-rich populations, 
while the bluer dwarf ellipticals contain a wider range of ages and lower 
metallicities than their more massive counterparts.  Radial SBF gradients reveal 
that IR color gradients are largely an age effect; the bluer dwarfs typically have
their youngest populations near their centers, while the redder giant ellipticals 
show only weak trends and in the opposite sense. Because of the population 
variations among bluer galaxies, distance measurements in the near-IR are best 
limited to red early-type galaxies.  We conclude with some practical guidelines for 
using WFC3/IR to measure reliable SBF distances.
\end{abstract}


\section{Introduction}

\subsection{Distance Measurements in the Big Picture}

Accurate distance measurement is central to both astrophysics
and cosmology.  Reliable distances are needed to convert observed properties of
galaxies (fluxes and angular sizes) into absolute physical quantities such as
luminosities, masses, ages, star formation rates, and dynamical time scales. 
In the local universe where peculiar velocities are significant, the distance
estimate is often a major source of uncertainty on these physical properties.
For instance, a recent review article on supermassive black holes (Kormendy \&
Ho 2013) notes that for many galaxies, the errors in the central black hole 
masses are dominated by the uncertainty in distance; yet, many authors 
neglect to include this important contribution to the uncertainty.

In the field of cosmology, the acceleration of the cosmic expansion was first
revealed by accurate distance measurements of Type~Ia supernovae (Riess et
al.\ 1998; Perlmutter et al.\ 1999).  The supernova distance estimates, combined
with flatness constraints for the universe 
provided by the cosmic microwave background (CMB)
power spectrum, are primarily responsible for inaugurating a new era of
``precision cosmology,'' central to which is the conclusion that the mass-energy
budget of the universe is dominated by ``dark energy.''  Now, with the exquisite
constraints on the CMB power spectrum at $z{\sim}1100$ (e.g., Bennett et 
al.\ 2013; Planck Collaboration 2015) and measurements of the 
baryon acoustic oscillation (BAO) 
scale at intermediate redshift (e.g., Eisenstein et al.\ 2005; Blake et
al.\ 2011), there has been a new effort to determine the local value of the
Hubble constant \ho\ to a precision of 1\% (Riess et al.\ 2011; Freedman et
al.\ 2012). As discussed in the foregoing references, this level of precision
is required for firm simultaneous constraints on cosmic geometry,
the dark energy equation of state, and the number of neutrino species.

Of course, the value of \ho\ has been controversial for decades, primarily owing
to systematic calibration errors (e.g., Freedman \& Madore 2010).
Determining \ho\ with a total uncertainty of no more than 1\% remains beyond 
the ability of any single distance measurement technique.
In order to achieve the required level of precision, it is helpful to have
multiple high-precision distance indicators to provide robust constraints on the
contributions from systematic errors.

The surface brightness fluctuation (SBF) method provides a measurement of the
mean brightness of the red giant branch stars in an early-type galaxy even 
though individual stars cannot be resolved (Tonry \& Schneider 1988). It was
introduced as a way to estimate distances with ${\sim}10$\% uncertainty out to
about 20~Mpc from ground-based astronomical images (see Tonry et al.\ 1990).
More recent applications have shown substantial improvements in both the
precision of the method and depth to which it can be applied (see the
reviews by Blakeslee 2012 and Fritz 2012), 
so that it has become one of a small number 
of methods capable of making a significant contribution to the problem
of constraining \ho\ to~1\%.  In the following sections, we discuss these
recent developments with the method and the need for a new calibration
at near-IR wavelengths.

\subsection{The Key Role of HST}

The latest revolution in our knowledge of the extragalactic distance scale
(decreasing the uncertainty from nearly a factor of two to less than 10\%)
has resulted primarily from observations made with the \textit{Hubble Space Telescope}
(\hst).  Starting in the mid-1990s, \hst\ was used to measure light curves for
samples of Cepheid variable stars in late-type galaxies out to ${\sim}20$~Mpc,
mostly as part of the Key Project on the Distance Scale
(e.g., Freedman et al.\ 1994; Ferrarese et al.\ 1996; Kelson et al.\ 1996; 
Saha et al.\ 1996, 1997; Silbermann et al.\ 1999).  The resulting Cepheid
distance estimates were used to calibrate various secondary distance indicators
and thereby derive the value of the Hubble constant \ho\
(Ferrarese et al.\ 2000; Gibson et al.\ 2000; Sakai et al.\ 2000; Mould et al.\ 2000).
The resulting value of \ho\ from the Key Project was $72\pm8$~\kmsmpc\ 
(Freedman et al.\ 2001) where the total uncertainty includes both 
random and systematic contributions. 
More recent Cepheid-based estimates are 
very similar to this value, but with reduced uncertainties of 3 to 4\%
(Riess et al.\ 2011; Freedman et al.\ 2012; Sorce et al.\ 2012).

The excellent angular resolution and photometric stability of \hst\ has also made it
possible to measure SBF with far better precision and to much larger distances
than was possible from the ground.  For instance, Jensen et
al.\ (2001) calibrated the SBF method for the near-IR F160W bandpass of the 
\hbox{NICMOS} NIC2 camera on \hst\ (Thompson et al.\ 1999)
and measured SBF distances to 16 galaxies beyond 40~Mpc,
including the first SBF distances reaching beyond 100~Mpc. 
Stellar population effects on the NICMOS F160W SBF magnitudes were explored in
detail by Jensen et al.\ (2003) using a larger sample of 65 galaxies.
In general, the SBF absolute magnitude in a given bandpass depends on stellar
population and must be calibrated using a population indicator, typically a
broadband color.

Following the installation of the Advanced Camera for Surveys (ACS) on \hst,
Mei et al.\ (2005b) produced the first calibration of the SBF method for the 
ACS Wide Field Channel (ACS/WFC) using F475W (\gacs) and F850LP (\zacs) data
for 84 galaxies from
the ACS Virgo Cluster Survey (ACSVCS; C\^ot\'e et al.\ 2004). In that work, 
the \zacs\ SBF measurements were calibrated for stellar population 
variations based on the observed \gz\ color; the resulting distances
enabled the first clear resolution of the depth of the Virgo cluster and provided
constraints on its triaxial structure (Mei et al.\ 2007).
In other studies based on ACS/WFC observations,
Cantiello et al.\ (2005, 2007) measured
multi-band SBF and color gradients in 21 galaxies, and Biscardi et al.\ (2008)
made the first optical SBF measurements beyond 100~Mpc.
As part of the ACS Fornax Cluster Survey (ACSFCS; Jord{\'a}n et al.\ 2007),
Blakeslee et al.\ (2009) refined the SBF calibration for the ACS F850LP bandpass
and determined the relative distance of the Virgo and Fornax clusters to a
precision of 1.7\%.  Additional ACS/WFC SBF measurements and a new calibration for 
the F814W bandpass were published by Blakeslee et al.\ (2010).
The launch of Wide Field Camera~3 with its powerful IR channel (WFC3/IR)
has greatly increased the distance to which SBF measurements can be made within
a single \hst\ orbit. However, the method must first be calibrated for selected
WFC3/IR passbands as done previously for NICMOS and ACS;
this is the primary goal of the present work.

\subsection{SBF Measurements in the Infrared}
The development of new infrared detectors in the 1990s allowed researchers to
successfully apply the SBF techniques to IR images for the first time 
(Luppino \& Tonry 1993; Pahre \& Mould 1994; Jensen, Luppino, \& Tonry 1996).
Because the SBF signal is dominated by the most luminous stars in a population,
and these tend to be quite red for evolved galaxies, SBF magnitudes are much
brighter in the~near-IR than at optical wavelengths. Additionally,
extinction by dust (both in our Galaxy and in the target galaxy) is much lower
at near-IR wavelengths. Depending on how it is distributed, dust can either
reduce the fluctuation signal (as for a uniform screen of foreground dust in 
the Galaxy), or, more commonly, bias the fluctuation signal towards higher
amplitudes and shorter SBF distances, as would occur if dusty regions were
clumpy on scales comparable to the size of the point-spread function (PSF). 
Clumpy dust is often associated with recent star formation, and bright young 
stars seriously bias SBF measurements as well.
The contrast between the fluctuations and
other point-like sources (globular clusters and background galaxies) is also
higher in the near-IR bands, reducing yet another source of uncertainty in the
SBF measurement. While the background in the IR is higher than at optical
wavelengths, the increased strength of the SBF signal more than compensates,
especially from space, where the thermal background at 1.1 and 1.6~\micron\ is
not significant.  The benefit of the much lower background, combined with the
excellent image quality and a very stable PSF, usually makes
near-IR SBF measurements with \hst\ much more accurate than measurements from
ground-based facilities.

However, the calibration of the SBF magnitudes as a function of stellar
population is potentially more complicated in the near-IR.  As noted
above, the trend of SBF magnitude with galaxy color is used for calibrating the
SBF distance measurements, both at optical and IR wavelengths.  At optical
wavelengths, the effects of age and metallicity variations on the SBF
calibration relations are largely degenerate, but this degeneracy begins to
break down in near-IR, and this can reveal interesting differences in the
stellar populations of galaxies.  Bluer elliptical and S0 galaxies typically
show signs of intermediate-age populations, and the asymptotic giant branch
(AGB) stars associated with those populations produce brighter fluctuations
(Jensen et al.\ 2003; Mieske, Hilker, \& Infante 2003, 2006).  
The stellar population variations may therefore produce
more scatter in IR SBF distance calibration, but the brightness of the
fluctuations at these wavelengths make them measurable to much larger distances;
thus, it is worth characterizing the behavior and limits of the calibration as
well as possible.

In this paper we report the results of a study to calibrate the IR SBF
distance measurement technique using new SBF measurements 
in the F110W (\Jir) and F160W (\Hir) bandpasses of WFC3/IR on \hst.
The calibration sample includes 16 galaxies 
spanning a wide range in galaxy luminosity and color.
The following section describes the observations and sample in more detail.
Section~3 discusses the data reductions and SBF measurements.
New SBF calibrations in \Jir\ and \Hir\ are presented in Section~4, 
while implications of the SBF measurements for the galaxy stellar populations are
discussed in Section~5. We provide our recommendations for measuring SBF
distances with WFC3/IR in Section~6, before concluding with a summary.

\section{WFC3/IR Observations}

In order to calibrate the SBF method for WFC3/IR, we selected
16 early-type galaxies, eight in each of the Virgo and
Fornax clusters, that already had high-quality ACS SBF measurements 
in \zacs\ and \gzacs\ colors (Blakeslee et al.\ 2009).
Table~\ref{tab:props} lists the properties of the galaxies that were targeted
in \hst\ program GO-11712 (PI: J.~Blakeslee).
The galaxies were chosen to cover the full color range of the ACSVCS and ACSFCS samples
so that the resulting calibration would be as generally applicable as possible.
Moreover, the NICMOS SBF calibration exhibited increased scatter for bluer galaxies
(Jensen et al.\ 2003), as have ground-based $I$-band SBF measurements 
(Mieske et al.\ 2006); exploring a broad color range should help us understand where
the WFC3/IR calibration becomes less reliable.

\begin{deluxetable*}{lccrrrrrc}
\tablecaption{Galaxy Properties}
 \tablewidth{0pt}
\tablehead{
\colhead{Galaxy} &
\colhead{Cluster\tablenotemark{a}} &
\colhead{Type\tablenotemark{b}} & 
\colhead{$m_B$\tablenotemark{c}} & 
\colhead{$R_e$\tablenotemark{d}}  &
\colhead{$M_B$\tablenotemark{e}} &
\colhead{$R_e$\tablenotemark{f}} &
\colhead{$A_J$\tablenotemark{g}} &
\colhead{Alt ID\tablenotemark{h}} \\
& & &
\colhead{(mag)} &
\colhead{(arcsec)} &
\colhead{(mag)} &
\colhead{(kpc)} &
\colhead{(mag)} &}
\startdata
 IC 1919 &  F & dS0 & 13.5 &  21.2 & $-$18.1 &  2.06 & 0.010 &   FCC 43 \\
 IC 2006 &  F &  E1 & 12.2 &  18.9 & $-$19.4 &  1.84 & 0.008 &   \dots  \\
NGC 1344 &  F &  E5 & 11.3 &  33.2 & $-$20.3 &  3.22 & 0.013 & NGC 1340 \\
NGC 1374 &  F &  E0 & 11.9 &  29.2 & $-$19.7 &  2.83 & 0.010 &  FCC 147 \\
NGC 1375 &  F &  S0 & 13.6 &  13.2 & $-$18.0 &  1.28 & 0.010 &  FCC 148 \\
NGC 1380 &  F &  S0 & 11.3 &  30.3 & $-$20.3 &  2.94 & 0.012 &  FCC 167 \\
NGC 1399 &  F &  E0 & 10.6 & 114.6 & $-$21.0 & 11.11 & 0.009 &  FCC 213 \\
NGC 1404 &  F &  E2 & 10.9 &  22.9 & $-$20.7 &  2.22 & 0.008 &  FCC 219 \\
\\[-7pt]
 IC 3025 &  V & dS0 & 14.8 &   9.8 & $-$16.4 &  0.78 & 0.015 &   VCC 21 \\
 IC 3032 &  V & dE2 & 14.7 &   9.1 & $-$16.6 &  0.73 & 0.026 &   VCC 33 \\
 IC 3487 &  V & dE6 & 14.8 &   9.3 & $-$16.4 &  0.74 & 0.015 & VCC 1488 \\
 IC 3586 &  V & dS0 & 14.5 &  20.2 & $-$16.7 &  1.61 & 0.032 & VCC 1695 \\
NGC 4458 &  V &  E1 & 12.9 &  20.4 & $-$18.2 &  1.63 & 0.017 & VCC 1146 \\
NGC 4472 &  V &  E2 &  9.3 & 210.8 & $-$21.9 & 16.87 & 0.016 & VCC 1226 \\
NGC 4489 &  V &  S0 & 12.8 &  37.1 & $-$18.4 &  2.97 & 0.020 & VCC 1321 \\
NGC 4649 &  V &  E2 &  9.8 &  98.1 & $-$21.4 &  7.85 & 0.019 & VCC 1978 
\smallskip
\enddata
\label{tab:props}
\tablenotetext{a}{Cluster: $V$ for Virgo and $F$ for Fornax.}
\tablenotetext{b}{Morphological type from the ACSVCS (C\^ot\'e et al.\ 2004)
                  and ACSFCS (Jord{\'a}n et al.\ 2007).}
\tablenotetext{c}{Apparent $B$-band magnitude (Vega).} 
\tablenotetext{d}{Effective radius in arcseconds, determined from the ACS and/or 
SDSS imaging (Ferrarese et al.\ 2006; Chen et al.\ 2010; P.\,C{\^o}t{\'e}, 
priv.~comm.).}  
\tablenotetext{e}{Absolute $B$ magnitude (Vega), corrected for Galactic extinction, 
and assuming distances of 16.5 and 20.0 Mpc for galaxies in Virgo and Fornax, 
respectively.}
\tablenotetext{f}{Effective radius in kpc, assuming same Virgo and Fornax distances 
as above.}
\tablenotetext{g}{Galactic extinction (Vega mag) in $J$-band, from Schlafly \& 
Finkbeiner (2011).}
\tablenotetext{h}{Alternate names from Virgo and Fornax Cluster Catalogs (Binggeli 
et al.\ 1985; Ferguson 1989), or alternate NGC designation in the case of NGC\,1344.}
\end{deluxetable*}

Each of these 16 galaxies were observed for one orbit,
split approximately equally between the \Jir\ and \Hir\ filters,
with four dithered exposures in each filter and total exposure times
varying with target visibility.  The same dither pattern was used for
each galaxy.

For this study we also downloaded archival WFC3/IR \Hir\ data for NGC~4258 
(GO-11570, PI: A.~Riess) and NGC~1316 (GO-11691, PI: P. Goudfrooij).
NGC~4258 is a late-type galaxy with a H$_2$O masers in
Keplerian orbits around a central black hole, enabling a geometric estimate of the
distance (Greenhill et al.\ 1995; Miyoshi et al.\ 1995;  Herrnstein et al.\ 1999).
While not ideal for SBF analysis, the importance of this galaxy to the 
absolute calibration of the extragalactic distance scale makes it an important 
target worthy of a trial SBF measurement. 
NGC~1316 is an early-type S0 galaxy in the Fornax cluster with extensive dust 
and signs of recent merging. Although it is also a poor SBF candidate, it is a useful 
comparison galaxy for this study because it has hosted \emph{four} type Ia supernovae.  

Tables~\ref{jmeasurements} and~\ref{hmeasurements} list the exposure times and sky
brightnesses for the all the \Jir\ and \Hir\ observations used in this study.
Additional information in these tables is discussed in the following sections.

\section{SBF Measurements}

The spatial fluctuations in the surface brightness of a smoothly distributed
population of stars, as found in elliptical and lenticular galaxies, arise due 
to the Poisson statistics of the discrete stars making up the galaxy, even when 
the stars cannot be resolved or detected individually 
(Tonry \& Schneider 1988). 
The SBF amplitude scales inversely with the square of the distance: nearby galaxies 
appear ``bumpy'' compared to more distant galaxies, where more stars are sampled by 
each resolution element and the $\sqrt{N}$ variation between regions
is therefore a smaller fraction of the total number of stars.
The fluctuations, which are dominated by the most luminous stars in a 
population, are blurred by the PSF; additional contributions
to the fluctuation signal arise from clumpy dust, globular clusters, background
galaxies, and foreground stars. 
The process of making an SBF measurement consists of
extracting and fitting the spatial Fourier power spectrum of the stellar 
fluctuations convolved with the PSF power spectrum 
and removing the contributions from extraneous sources. 
The resulting fluctuation power is used to compute the fluctuation magnitude. 

Procedures for measuring surface brightness fluctuations have been described in
detail by several authors (e.g., Tonry et al.\ 1990, 1997; 
Blakeslee et al.\ 1997; Jensen et al.\ 1998; Mei et al.\ 2005a; Fritz 2012). 
The description here provides a concise
overview of the process steps that are either unique to this study or are
particularly relevant to the WFC3/IR SBF measurements.

\subsection{Data Reduction\label{datareduction}}
 
The first step in the SBF data reduction process involves producing a 
calibrated, combined, and background-subtracted image ready for further SBF 
analysis. We used the images from the \hst\ archive reduced using the standard 
pipeline through the flat-fielding stage (\emph{flt} files). 
From that point, we adopted a reduction procedure that 
differs from the standard pipeline. 

We combined the individual flat-fielded exposures using integer pixel 
shifts after fitting each image for background and identifying cosmic rays;
in order to avoid introducing
correlated noise between pixels, fractional pixel registration was \emph{not} used. 
Using integer pixel shifts results in slightly lower spatial resolution in 
the combined image, but preserves the independence of noise from pixel to 
pixel, which is important for fitting the SBF power spectrum. 

\begin{figure*}
\center
\includegraphics[width=5in]{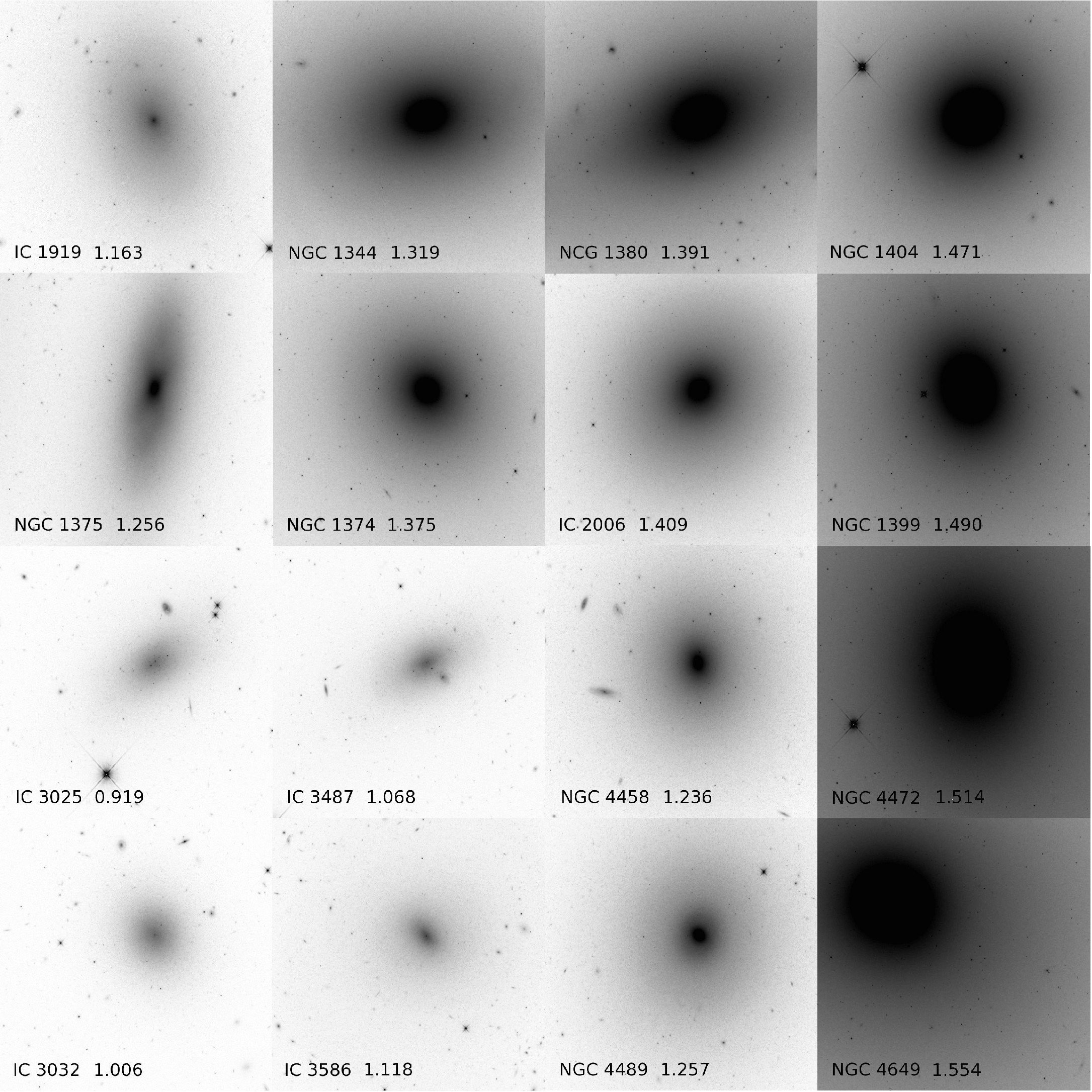}
\caption{Fornax (top two rows) and Virgo cluster galaxies (bottom two rows). 
Each image is displayed with background subtracted and using the same 
upper and lower limits and logarithmic stretch, within the same field of view 
100 arcsec on a side. The bluer galaxies are on the left and the redder
giant ellipticals are on the right. Measured values of \gz\ (AB) are
shown next to the galaxy labels.\\
\label{imagefigure}}
\end{figure*}

In a second difference from the standard pipeline reduction, the clean 
combined images were \emph{not} corrected for the WFC3/IR geometrical 
distortion nor combined using \emph{astrodrizzle}. 
Our analysis therefore includes the $\sim$10\% difference in plate scale
between the $x$ and $y$ axes, causing our images to appear somewhat narrower
horizontally than they do on the sky (Fig.~\ref{imagefigure}). 
The SBF procedure involves taking the spatial Fourier power spectrum, and
correlated noise between pixels resulting from fractional pixel shifts
and interpolated pixel values when correcting for geometrical distortion
can produce a slope in the white noise component of the power spectrum.

Finally, we chose not to apply the correction to pixel size in the
$y$-axis of WFC3/IR images. The pixel map correction usually used for WFC3/IR
corrects for PSF variations but is inappropriate for extended objects. 
The WFC3/IR focal plane is tilted, and the size (area) of the pixels on the
sky varies by ${\sim}8$\% from the center to the upper and lower edges of
the frame (Kalirai et al.\ 2010).
When the data are divided by the flat field image (as is done for the 
\emph{flt} files in the \hst\ archive), the varying pixel sensitivity is 
removed, effectively making the pixels equally sensitive to uniform 
illumination.
Flattening the images in this way creates a variation in sensitivity for point sources
from the center to the top and bottom edges that is usually corrected in the
pipeline data reduction process using \emph{astrodrizzle}.
Since we are interested in accurately measuring the surface brightnesses of
the galaxies and avoiding correlated noise between pixels, we chose not to
correct for pixel size variation in our SBF data reduction process.
The practical effect of this decision is that the PSF at the extreme
upper and lower regions of the image does not match the PSF near the center.
We have carefully chosen PSF reference stars from the same vertical region 
of the field of view as the galaxies being analyzed (usually very near the 
center) to avoid any systematic offset between the PSF photometric 
normalization or power spectrum shape and the galaxy fluctuations. 
The chosen PSF stars were all unresolved, isolated from other bright objects,
and much brighter than the globular cluster population. 

\begin{deluxetable*}{lccccccc}
\tablecaption{\Jir\ SBF Measurements\label{jmeasurements}}
\tablewidth{0pt}
\tablehead{
\colhead{Galaxy} &
\colhead{Exposure} &
\colhead{Background} &
\colhead{Annulus} &
\colhead{$\left<{\rm gal/sky}\right>$\tablenotemark{a}} &
\colhead{gal/sky} &
\colhead{SBF $S/N$\,\tablenotemark{b}} &
\colhead{$\overline{m}_{110}$\,\tablenotemark{c}} \\
\colhead{} &
\colhead{(sec)} &
\colhead{(AB mag/arcsec$^2$)}&
\colhead{(arcsec)}&
\colhead{average} &
\colhead{range} &
\colhead{$(P_0{-}P_r)/P_1$} &
\colhead{(AB mag)}
}
\startdata
\emph{\ \ Fornax}&&&&&&&\\
IC 1919  & 997 & 21.74 & 8\,--\,33 & 1.3 & 0.9\,--\,2.6& 33 & $28.31\pm0.05$ \\
IC 2006  & 997 & 22.40 & 4\,--\,33 & 12  & 6\,--\,58   & 43 & $28.63\pm0.05$ \\
NGC 1344 & 997 & 22.10 & 4\,--\,33 & 19  & 11\,--\,85  & 83 & $28.37\pm0.05$ \\
NGC 1374 & 997 & 22.02 & 4\,--\,33 & 10  & 5\,--\,48   & 47 & $28.55\pm0.06$ \\
NGC 1375 &1197 & 22.64 & 4\,--\,33 & 5.8 & 3.1\,--\,27 & 59 & $28.19\pm0.05$ \\
NGC 1380 &1197 & 22.22 & 6\,--\,33 & 29  & 19\,--\,124 & 49 & $28.51\pm0.04$ \\
NGC 1399 &1197 & 22.15 & 4\,--\,33 & 41  & 23\,--\,194 & 52 & $28.79\pm0.05$ \\
NGC 1404 & 997 & 22.04 & 4\,--\,33 & 32  & 17\,--\,173 & 47 & $28.68\pm0.06$ \\
\\[-7pt]
\emph{\ \ Virgo}&&&&&&&\\
IC 3025  & 997 & 21.84 & 8\,--\,17 & 0.8 & 0.8         & 17 & $28.36\pm0.06$ \\
IC 3032  & 997 & 21.81 & 8\,--\,17 & 0.8 & 0.8         & 34 & $27.95\pm0.07$ \\
IC 3487  & 997 & 21.84 & 4\,--\,33 & 0.4 & 0.2\,--\,2.5& 27 & $28.16\pm0.10$ \\
IC 3586  & 997 & 22.17 & 4\,--\,33 & 0.8 & 0.5\,--\,3.6& 34 & $28.05\pm0.08$ \\
NGC 4458 & 997 & 22.12 & 4\,--\,67 & 4.4 & 0.5\,--\,22 & 47 & $28.01\pm0.05$ \\
NCG 4472 & 748 & 22.10 & 4\,--\,67 & 68  & 17\,--\,249 & 40 & $28.38\pm0.06$ \\
NGC 4489 & 997 & 22.23 & 8\,--\,67 & 4.5 & 0.6\,--\,8  & 58 & $27.81\pm0.06$ \\
NGC 4649 & 997 & 22.10 & 4\,--\,67 & 61  & 14\,--\,250 & 43 & $28.44\pm0.06$ \\
\\[-7pt]
\emph{\ \ Supernova host}&&&&&&&\\
NGC 1316 &1396 & 22.08 & 33\,--\,67 & 15  & 8\,--\,34 & 166 & $28.26\pm0.07$
\\[-7pt]
\enddata
\tablenotetext{a}{Weighted average ratio of the galaxy surface brightness to 
sky background surface brightness within the measurement region.}
\tablenotetext{b}{Weighted average ratio of the SBF fluctuation power to the 
white noise component $P_1$ of the spatial power spectrum.}
\tablenotetext{c}{SBF magnitudes have been corrected for Galactic extinction
using values from Schlafly \& Finkbeiner (2011).}
\end{deluxetable*}

\begin{deluxetable*}{lccccccc}
\tablecaption{\Hir\ SBF Measurements\label{hmeasurements}}
\tablewidth{0pt}
\tablehead{
\colhead{Galaxy} &
\colhead{Exposure} &
\colhead{Background} &
\colhead{Annulus} &
\colhead{$\left<{\rm gal/sky}\right>$\tablenotemark{a}} &
\colhead{gal/sky} &
\colhead{SBF $S/N$\,\tablenotemark{b}} &
\colhead{$\overline{m}_{160}$\,\tablenotemark{c}} \\
\colhead{} &
\colhead{(sec)} &
\colhead{(AB mag/arcsec$^2$)}&
\colhead{(arcsec)}&
\colhead{average} &
\colhead{range} &
\colhead{$(P_0{-}P_r)/P_1$} &
\colhead{(AB mag)}
}
\startdata
\emph{\ \ Fornax}&&&&&&&\\
IC 1919  & 997 & 21.90 & 8\,--\,33 & 1.8 & 1.3\,--\,3.8 & 43 & $27.55\pm0.06$ \\
IC 2006  & 997 & 22.13 & 4\,--\,33 & 12  & 6\,--\,59    & 59 & $27.86\pm0.04$ \\
NGC 1344 & 997 & 21.90 & 4\,--\,33 & 20  & 12\,--\,91   & 77 & $27.52\pm0.05$ \\
NGC 1374 & 748 & 21.95 & 4\,--\,33 & 12  & 6\,--\,59    & 45 & $27.88\pm0.05$ \\
NGC 1375 &1197 & 22.36 & 4\,--\,33 & 5.5 & 2.9\,--\,27  & 62 & $27.42\pm0.05$ \\
NGC 1380 &1197 & 21.95 & 6\,--\,67 & 29  & 19\,--\,120  & 71 & $27.80\pm0.05$ \\
NGC 1399 &1197 & 21.95 & 4\,--\,33 & 46  & 25\,--\,220  & 52 & $28.03\pm0.04$ \\
NGC 1404 & 997 & 22.07 & 4\,--\,33 & 43  & 22\,--\,235  & 50 & $27.94\pm0.06$ \\
\\[-7pt]
\emph{\ \ Virgo}&&&&&&&\\
IC 3025  & 997 & 21.76 & 8\,--\,17 & 0.3 & 0.3 & 27     & $27.72\pm0.07$ \\
IC 3032  & 997 & 21.86 & 8\,--\,17 & 1.0 & 1.0 & 17     & $26.94\pm0.09$ \\
IC 3487  & 997 & 21.75 & 4\,--\,33 & 0.4 & 0.2\,--\,2.7 & 27 & $27.42\pm0.09$ \\
IC 3586  & 997 & 22.01 & 4\,--\,33 & 0.9 & 0.5\,--\,3.8 & 38 & $27.26\pm0.04$ \\
NGC 4458 & 997 & 21.94 & 4\,--\,67 & 4.7 & 0.5\,--\,23  & 56 & $27.35\pm0.05$ \\
NCG 4472 & 748 & 21.83 & 4\,--\,67 & 71  & 17\,--\,258  & 47 & $27.59\pm0.05$ \\
NGC 4489 & 997 & 22.07 & 8\,--\,67 & 4.8 & 0.6\,--\,8   & 67 & $27.10\pm0.05$ \\
NGC 4649 & 997 & 21.90 & 4\,--\,67 & 68  & 15\,--\,283  & 67 & $27.63\pm0.05$ \\
\\[-7pt]
\emph{\ \ Supernova host}&&&&&&&\\
NGC 1316 &2796 & 21.85 & 33\,--\,67 & 21 & 9\,--\,36 & 111 & $27.35\pm0.07$\\
\\[-7pt]
\emph{\ \ Maser host}&&&&&&&\\
NGC 4258 &2012 & 21.91 & irreg. & 10  & 47 & 47 & $25.30\pm0.06$
\\[-7pt]
\enddata
\tablenotetext{a}{Weighted average ratio of the galaxy surface brightness to
sky background surface brightness within the measurement region.}
\tablenotetext{b}{Weighted average ratio of the SBF fluctuation power to the white noise
component of the spatial power spectrum.}
\tablenotetext{c}{SBF magnitudes have been corrected for Galactic extinction
using values from Schlafly \& Finkbeiner (2011). \\}
\end{deluxetable*}

\begin{figure*}
\center
\includegraphics[width=5in]{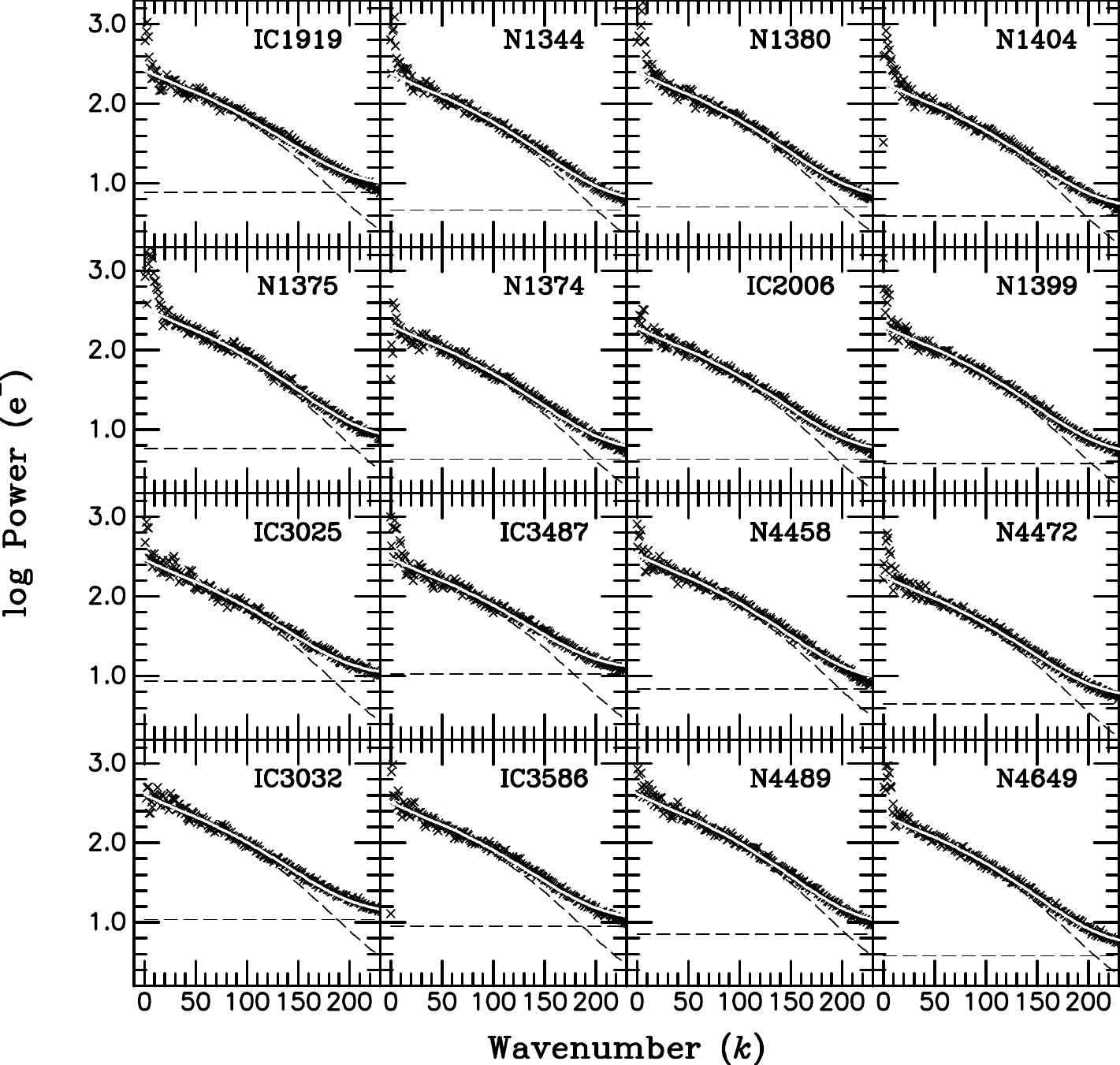}
\caption{Fits to the spatial power spectra for the 16 galaxies in Fornax 
and Virgo (superimposed white line). 
The dashed lines indicate the scaled PSF power spectrum and the white noise 
(flat) components.\\
\label{powerspectra}}
\end{figure*}

\subsection{Measuring Fluctuations}

SBF measurements are made by fitting the Fourier spatial power spectrum
of the stellar fluctuations in a given region of a galaxy with the normalized power
spectrum of the PSF.  There are several steps to produce the
spatial power spectrum: (i) the background level is estimated and 
subtracted; (ii) extraneous objects (globular clusters, background galaxies, 
and dusty regions) are identified, measured, and masked; (iii) a smooth isophotal
model is fitted to the galaxy profile and subtracted; (iv) isolated bright stars
are extracted and used to determine the power spectrum of the PSF; 
(v) the Fourier power spectra are computed for the 
stellar fluctuations, the masked galaxy profile, and the PSF; and 
(vi) the power spectrum of the data is fitted and normalized to determine 
the SBF power in units of flux, with the power from undetected globular clusters and
galaxies subtracted,  
from which a fluctuation magnitude is computed in the established way
(e.g., Tonry \& Schneider 1988; Jensen et al.\ 1998; Mei et al.\ 2005a).

It is important to accurately measure and subtract the IR background 
before computing fluctuation magnitudes. Estimating the background level
was done in an iterative process of cross-checking values measured in different ways.
For the small galaxies, we measured the flux in the corners of the frames. 
For the larger ellipticals, we also determined the background using the best 
fit to a $r^{1/4}$ profile for each galaxy. These estimates were then compared to a
measurement of the background made by iteratively computing a smooth model
for each galaxy and looking at the residual background in the field of view.
By adjusting the sky level offset, we optimized the galaxy models such
that the residual background was not systematically positive or negative.
The difference between the sky values determined using the different methods
was used as an estimate of the uncertainty in the sky level, and
the SBF analysis was repeated to determine the uncertainty
in fluctuation magnitude due to sky level uncertainty. The SBF magnitude is
normalized by the mean galaxy brightness at the location of the SBF measurement.
For the smaller galaxies with the lowest surface brightness 
(see Fig.~\ref{imagefigure}), the
background level was most accurately measured. For the largest galaxies that
extend beyond the limits of the field of view, the background was most difficult
to measure accurately, but its influence on the SBF measurement was also minimal.
The uncertainty in the SBF magnitude is therefore relatively insensitive to the
uncertainty in the background measurement.

Typical background levels at \Jir\ were 1.3 e$^-$s$^{-1}$pix$^{-1}$, or 
$22.10\,{\pm}\,0.23$ AB~mag\,arcsec$^{-2}$; 
at \Hir\ we measured 0.67 e$^-$s$^{-1}$pix$^{-1}$, or $21.95\,{\pm}\,0.15$ 
AB~mag\,arcsec$^{-2}$ (see Tables \ref{jmeasurements} and \ref{hmeasurements}).
These sky values were found to be consistent with published empirical 
measurements of the \hst\ IR background, which is dominated by scattered 
zodiacal light in the Solar System (Pirzkal 2014).
The \Jir\ background levels were also affected by 
the diffuse upper-atmosphere He emission line at 1.083 $\mu$m
(Brammer et al.\ 2014). 
While very few individual exposures were strongly affected, 
the residual background variation among the data is 
larger at \Jir\ than at \Hir. 
The background variability between observations of different galaxies 
(see Tables \ref{jmeasurements} and \ref{hmeasurements}) is not directly
related to the uncertainty in the background measurement for a particular galaxy,
which was determined independently for each galaxy.

After the background had been subtracted, we then identified and masked
any non-fluctuation point sources in the field of view. 
Background galaxies and globular clusters were identified and their 
brightnesses measured using the SExtractor software (Bertin \& Arnouts 1996)
and adopting aperture corrections taken from the WFC3 instrument
webpage.\footnote{http://www.stsci.edu/hst/wfc3/phot\_zp\_lbn}
We started by making an initial fit to the galaxy and subtracting it to
make it easier for SExtractor to identify and measure all the compact objects 
in the field (we used a noise model for SExtractor that accounts for the 
subtracted galaxy and prevents the software from confusing the stellar 
fluctuations with globular clusters). 
Using the SExtractor output, we then created a fit to the luminosity functions 
of the globular clusters and background galaxies.
For most of the galaxies, we used a Gaussian of width 1.2 mag to fit the 
luminosity functions; we found that a width of 1.35 mag fit better for 
NGC~1399 and NGC~4472. 
We adopted a luminosity function peak absolute magnitude of  
$M_V{=}-7.40$, $M_J{=}-8.26$, and $M_H{=}-8.29$
(e.g., Harris 2001; Frogel et al.\ 1978).
The background galaxies are assumed to follow a
power law distribution with power law slope of 0.25, which for most of our
observations results in a normalization of about one to two
galaxies per square arcsec at 34.50 mag AB (Retzlaff et al.\ 2010; 
Windhorst et al.\ 2011).
These fits allowed us to integrate the contribution to the SBF signal
fainter than the completeness limit and correct the final SBF magnitude
accordingly. A mask was then created from the list of objects that removed 
all objects brighter than the limiting cutoff magnitude.

The next step was to fit a final smooth model to the galaxy profile with 
the sky and point sources removed. We used an iterative procedure of fitting 
elliptical isophotes to the galaxy surface brightness profile, allowing
the procedure to adjust the centers, ellipticities, and orientations of
the elliptical apertures. This model galaxy was later used to normalize the
fluctuation signal. 

With the sky background and  mean galaxy surface brightness removed and 
contaminating point sources masked, we then computed the Fourier transform
and spatial power spectrum of the data in circular annuli 
(see Tables~\ref{jmeasurements} and \ref{hmeasurements} for the sizes of 
the annular regions).
The measurements were repeated using elliptical annuli for the six most
elongated galaxies with apparent radial gradients in their SBF magnitudes.
The power spectrum was normalized by the
mean galaxy luminosity so that the fluctuation amplitude should be the same
in each annulus. The purpose of measuring fluctuations in multiple annuli
was to look for consistency between regions with varying surface brightness,
globular cluster population, and distance from the center.
It also allowed us to measure the radial gradients in fluctuation
amplitude, and thus stellar population, as described in 
Section~\ref{radialgradients}.

The measured fluctuation power spectrum is a convolution of the pixel-to-pixel
variation in the number of stars and the PSF
of the instrument. We therefore require a robust measurement of the PSF
power spectrum to determine the fluctuation amplitude.
We extracted isolated bright stars from the central region of the 
detector field of view and computed their power spectra. 
Because PSF stars are not all uniformly centered on the pixels, there is
some variation that naturally arises in the PSF power spectra. 
To determine which PSF stars best fitted the power spectrum
for a particular galaxy,  
we combined the cleanest and brightest PSF stars from several observations,
and then repeated the SBF measurements using a variety of PSF stars.
We computed the uncertainty in SBF magnitude attributable to the PSF
variations by determining the range of plausible PSF fits based on the 
shape of the power spectra and the quality of the fits.
We then used a common composite PSF to measure SBF consistently in all the 
galaxies (the same dither pattern was used for all the observations). 
As an added check on the PSF uniformity and fit quality, we also fitted the 
observed power spectra to the ``Tiny Tim'' PSF
models\footnote{\url{http://tinytim.stsci.edu/cgi-bin/tinytimweb.cgi}} 
(Krist et al.\ 2011).
The Tiny Tim model PSF for each filter was convolved with Gaussians of 
various widths to construct a library of model PSFs to provide better 
matches to the data, which had been combined using integer pixel offsets
and no geometrical distortion corrections (see Sec.~\ref{datareduction}).
The SBF magnitudes computed using the Tiny Tim model PSFs were then 
compared to those derived from the combined empirical PSFs and the range
of plausible fits was used to determine the uncertainty due to PSF variations
(typically 0.04 mag).

Fluctuation magnitudes were computed by fitting the normalized PSF
power spectrum to the galaxy power spectrum 
$P(k) = P_0 E(k) + P_1$,
where $E(k)$ is the expectation power spectrum, which includes the smooth
galaxy profile and the combined annular region and external object mask, all
convolved
with the normalized PSF power spectrum (see Fig.~\ref{powerspectra} for 
power spectra and fit components). The fits excluded the lowest wavenumbers
$k\,{<}\,10$, which are affected by large-scale galaxy and sky subtraction errors.
$P_1$ is the white noise component, which is flat for uncorrelated 
pixel-to-pixel noise. 
The scale factor used to best match the data corresponds to the flux in
SBF power in units of e$^-$s$^{-1}$. It is then straight-forward to compute
the fluctuation magnitude:
$$ \overline{m} = -2.5\log(P_0-P_r) + M_1 $$
where $P_0$ is the fluctuation power and $P_r$ is the contribution from
point sources fainter than the completeness limit. $M_1$ is the zero
point for the filter+detector (26.8223 AB for \Jir\ and 
25.9463 AB for \Hir). The AB magnitude is 0.7595~mag larger than the Vega
equivalent at \Jir, and 1.2514~mag larger than the Vega magnitude at \Hir.
AB magnitudes and colors were corrected for Galactic extinction using
the values published by Schlafly \& Finkbeiner (2011). 
The \gz\ colors from Blakeslee et al.\ (2009) were adjusted to make them 
consistent with the Schlafly \& Finkbeiner extinction values.
Individual sky background levels, exposure times, and SBF $S/N$ values are 
listed in Tables~\ref{jmeasurements} and \ref{hmeasurements}.

Absolute fluctuation magnitudes \Mbarj\ and \Mbarh\ (Table~\ref{absmags})
were then computed 
using both individual distance modulus measurements and average 
cluster distances of $(m{-}M)=$31.51 mag for the Fornax cluster and 31.09 
for Virgo (all from the optical ACS SBF measurements of Blakeslee et al.\ 2009). 
The radial extent of each cluster (0.053 mag 
for Fornax and 0.085 mag for Virgo) was adopted as the uncertainty 
on the average cluster distance moduli.
The colors originally reported by Blakeslee et al.\ (2009) were
recomputed for the apertures used in this study (circular and elliptical).

There are several sources of uncertainty in the SBF measurement procedure
that we quantified by exploring the range of input parameters, as we did
for the uncertainty due to the PSF fit. Average values (and ranges) of the 
uncertainties we measured and adopted for the final SBF measurements are 
listed in Table~\ref{uncertainties}.
Not all sources of uncertainty are 
completely independent---residual errors in sky subtraction can affect the
galaxy or PSF fit, for example---so the total uncertainties listed in
Table~\ref{absmags} are not a simple quadrature addition of all sources 
listed in Table~\ref{uncertainties}; we estimated the fraction of the
power spectrum fit ($P_0$) uncertainty that results from the PSF fit and sky 
subtraction 
separately before adding all independent sources of error in quadrature.
The distance modulus uncertainties were
included in the individual-distance values of \Mbar. The cluster-distance
\Mbar\ values include the cluster distance dispersion values from
Blakeslee et al.\ (2009) in the total uncertainty.  
 
\begin{deluxetable*}{lccccccc}
\tablecaption{Absolute SBF Magnitudes\label{absmags}}
\tablewidth{0pt}
\tabletypesize{\footnotesize}
\tablehead{
\colhead{Galaxy} & 
\colhead{\mM\tablenotemark{a}} &
\colhead{\gz \tablenotemark{b}} &
\colhead{\JH} & 
\colhead{\Mbarj \tablenotemark{c}} & 
\colhead{\Mbarj \tablenotemark{d}} &
\colhead{\Mbarh \tablenotemark{c}} & 
\colhead{\Mbarh \tablenotemark{d}}
}
\startdata
\emph{\ \ Fornax}&&&&&&&\\
IC 1919  & $31.485\pm0.073$ & $1.163\pm0.037$ & $0.223\pm0.007$ & $-3.18\pm0.09$ & $-3.20\pm0.08$ & $-3.94\pm0.09$ & $-3.96\pm0.08$\\
IC 2006  & $31.525\pm0.086$ & $1.409\pm0.013$ & $0.263\pm0.013$ & $-2.90\pm0.10$ & $-2.88\pm0.07$ & $-3.67\pm0.10$ & $-3.65\pm0.07$\\
NGC 1344 & $31.603\pm0.068$ & $1.319\pm0.007$ & $0.260\pm0.008$ & $-3.23\pm0.08$ & $-3.14\pm0.07$ & $-4.08\pm0.08$ & $-3.99\pm0.07$\\
NGC 1374 & $31.458\pm0.070$ & $1.375\pm0.011$ & $0.251\pm0.014$ & $-2.91\pm0.09$ & $-2.96\pm0.08$ & $-3.58\pm0.09$ & $-3.63\pm0.07$\\
NGC 1375 & $31.500\pm0.072$ & $1.256\pm0.029$ & $0.224\pm0.016$ & $-3.31\pm0.09$ & $-3.32\pm0.07$ & $-4.08\pm0.09$ & $-4.09\pm0.07$\\
NGC 1380 & $31.632\pm0.075$ & $1.391\pm0.007$ & $0.286\pm0.006$ & $-3.12\pm0.09$ & $-3.00\pm0.07$ & $-3.83\pm0.09$ & $-3.71\pm0.07$\\
NGC 1399 & $31.596\pm0.091$ & $1.490\pm0.005$ & $0.302\pm0.012$ & $-2.81\pm0.10$ & $-2.72\pm0.07$ & $-3.57\pm0.10$ & $-3.48\pm0.07$\\
NGC 1404 & $31.526\pm0.072$ & $1.471\pm0.006$ & $0.292\pm0.005$ & $-2.85\pm0.09$ & $-2.83\pm0.08$ & $-3.59\pm0.10$ & $-3.57\pm0.08$\\
\\[-7pt]
\emph{\ \ Virgo}&&&&&&&\\
IC 3025  & $31.421\pm0.130$ & $0.919\pm0.074$ & $0.183\pm0.019$ & $-3.06\pm0.14$ & $-2.73\pm0.11$ & $-3.70\pm0.15$ & $-3.37\pm0.11$\\
IC 3032  & $30.886\pm0.133$ & $1.006\pm0.030$ & $0.184\pm0.036$ & $-2.94\pm0.15$ & $-3.14\pm0.11$ & $-3.95\pm0.16$ & $-4.15\pm0.13$\\
IC 3487  & $31.053\pm0.134$ & $1.068\pm0.060$ & $0.132\pm0.023$ & $-2.89\pm0.17$ & $-2.93\pm0.13$ & $-3.63\pm0.16$ & $-3.67\pm0.13$\\
IC 3586  & $31.093\pm0.080$ & $1.118\pm0.040$ & $0.188\pm0.007$ & $-3.04\pm0.11$ & $-3.04\pm0.11$ & $-3.83\pm0.09$ & $-3.83\pm0.13$\\
NGC 4458 & $31.063\pm0.070$ & $1.236\pm0.049$ & $0.237\pm0.026$ & $-3.05\pm0.08$ & $-3.08\pm0.10$ & $-3.71\pm0.09$ & $-3.74\pm0.10$\\
NCG 4472 & $31.116\pm0.075$ & $1.514\pm0.006$ & $0.291\pm0.013$ & $-2.74\pm0.10$ & $-2.71\pm0.10$ & $-3.53\pm0.09$ & $-3.50\pm0.10$\\
NGC 4489 & $30.935\pm0.069$ & $1.257\pm0.014$ & $0.226\pm0.013$ & $-3.13\pm0.09$ & $-3.28\pm0.10$ & $-3.84\pm0.08$ & $-3.99\pm0.10$\\
NGC 4649 & $31.082\pm0.079$ & $1.554\pm0.006$ & $0.311\pm0.013$ & $-2.64\pm0.10$ & $-2.65\pm0.11$ & $-3.45\pm0.09$ & $-3.46\pm0.10$\\
\\[-7pt]
\emph{\ \ Supernova host}&&&&&&&\\
NGC 1316 & $31.606\pm0.065$ & $1.374\pm0.007$ & $0.272\pm0.007$ & $-3.35\pm0.10$ & $-3.25\pm0.09$ & $-4.26\pm0.10$ & $-4.16\pm0.09$\\
\\[-7pt]
\emph{\ \ Maser host}&&&&&&&\\
NGC 4258 & $29.404\pm0.048$ & $1.361\pm0.044$ & \nodata         & \nodata        & \nodata        & $-4.10\pm0.08$ & \nodata
\enddata
\tablecomments{All magnitudes are on the AB system and extinction corrected.}
\tablenotetext{a}{Blakeslee et al.\ (2009) except NGC~4258, Humphreys et al.\ (2013).} 
\tablenotetext{b}{Galaxy colors from Blakeslee et al.\ (2009) have been 
updated to match the apertures used in this study, and corrected for 
extinction using Schlafly \& Finkbeiner (2011).}
\tablenotetext{c}{\Mbar\ computed using the individual distance moduli shown in the second column.}
\tablenotetext{d}{\Mbar\ computed using average cluster distances of 31.51 mag for Fornax and 31.09 for Virgo. 
Uncertainties on \Mbar\ include the cluster depths of 0.053 mag (Fornax) and 0.085 mag (Virgo).} 
\end{deluxetable*}

\begin{deluxetable}{lcc}
\tablecaption{Average Uncertainties\label{uncertainties}}
\tablewidth{0pt}
\tablehead{
\colhead{Source} &
\colhead{$\sigma$ (mag)} &
\colhead{Range (mag)} 
}
\startdata
Power spectrum fit $P_0$& 0.06 & 0.04\,--\,0.10 \\
PSF fit & 0.04 & 0.026\,--\,0.067 \\
Background subtraction & 0.01 & 0.002\,--\,0.062 \\
Galaxy subtraction & 0.007 & 0.001\,--\,0.022\\
\gz\ color uncertainty & 0.025 & 0.005\,--\,0.074 \\
\JH\ color uncertainty & 0.013 & 0.002\,--\,0.036 \\
Distance modulus (individual) & 0.086 & 0.068\,--\,0.134 \\
Distance modulus (Virgo) & 0.085 & 0.085 \\
Distance modulus (Fornax) & 0.053 & 0.053 \\
Total statistical \Mbar\ uncertainty & 0.10 & 0.08\,--\,0.17
\enddata
\\
\end{deluxetable}

\section{Analysis}
\subsection{Calibration of the WFC3/IR SBF Distance Scale}

The SBF signal can be detected with \hst\ in modest \Jir\ and \Hir\ exposures (an
orbit or less) out to $\sim$100~Mpc (e.g., Jensen et al.\ 2001).
To take full advantage of WFC3/IR observations of early-type
galaxies collected for a variety of purposes and measure accurate distances,
we calibrated the WFC3/IR \Jir\ and \Hir\ SBF distances by fitting the 
absolute fluctuation magnitudes \Mbar\ 
as a function of both optical \gz\ and IR \JH\ colors.
Determining the \Mbar\ values requires us to adopt
a distance modulus \mM\ for each galaxy. We used the \zacs\ SBF distance
moduli measured using ACS (Blakeslee et al.\ 2009).
These measurements provide a consistent distance reference 
accurate to $\lesssim\,$0.08 mag for most of the galaxies,
although the calibration becomes systematically less certain for $\gz<1.05$~mag.
We also adopted \gz\ values from the same ACS data, 
supplementing the published values from Blakeslee et al.\ (2009) with updated 
color measurements made using the original images in annuli that  matched 
our IR observations.
The ACS colors and distance moduli used to calibrate the IR SBF distance
scale are listed in Table~\ref{absmags}. Extinction-corrected \JH\
colors were measured using our WFC3/IR images.

\begin{figure*}
\center
\epsscale{1.0}
\plottwo{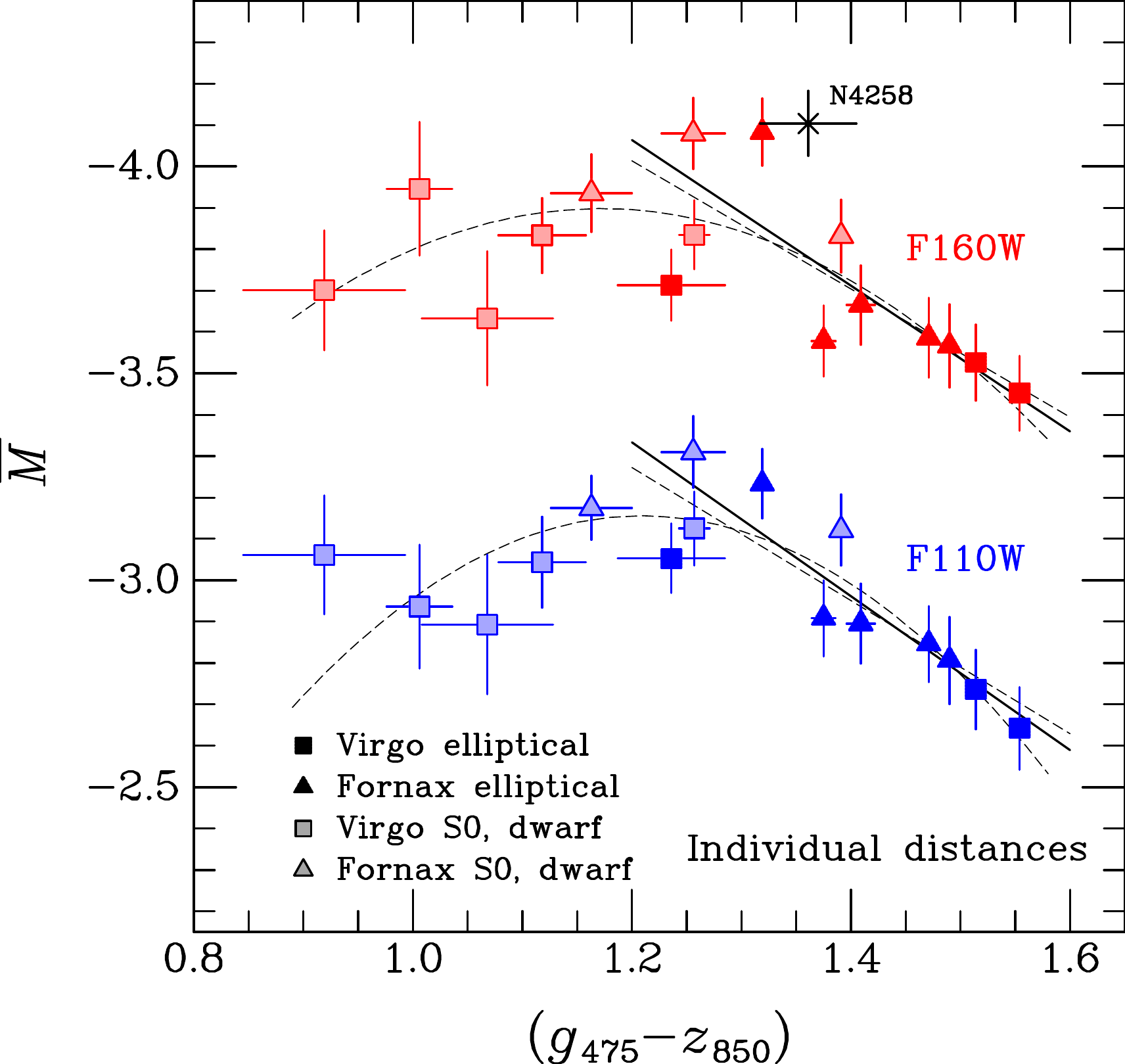}{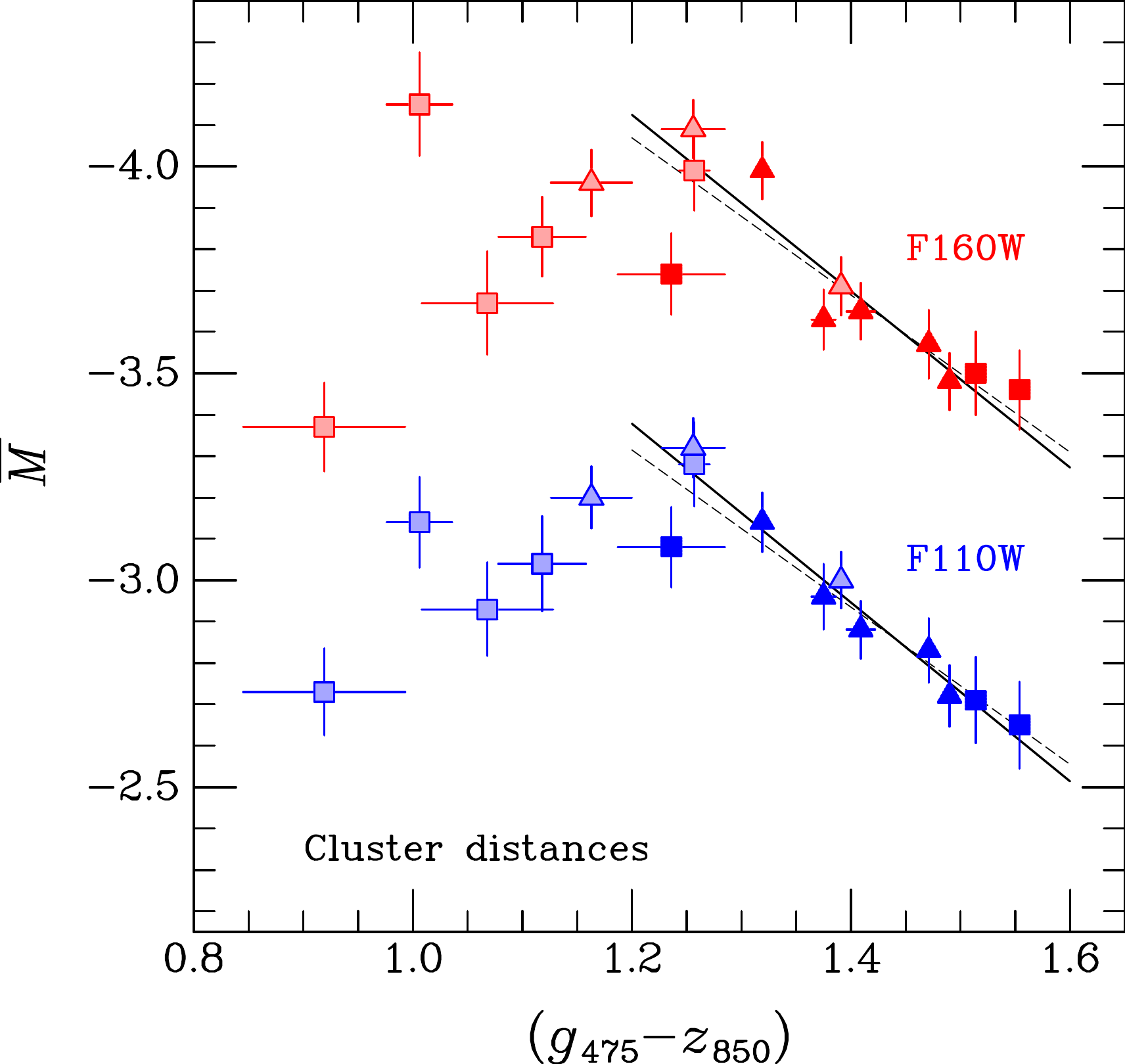}
\caption{Fits to the absolute fluctuation magnitudes \Mbar\ as a
function of \gz\ color.
We also plot \Mbar\ for NGC~4258 for comparison. 
The calibration shown on the left is based on distances 
derived using the individual $z$-band SBF distances. The panel on the 
right shows the calibration using average cluster distances for Virgo and
Fornax. Red points indicate \Hir\ measurements (top set) and blue symbols 
are the \Jir\ measurements (lower set). The dashed calibration lines 
show the quadratic fit and the linear calibration including IC~1919; 
the color of IC~1919 is intermediate between the other blue dwarf
galaxies and the elliptical galaxies in our sample.\\ 
\label{calibgz}}
\end{figure*}

\begin{figure*}
\center
\epsscale{1}
\plottwo{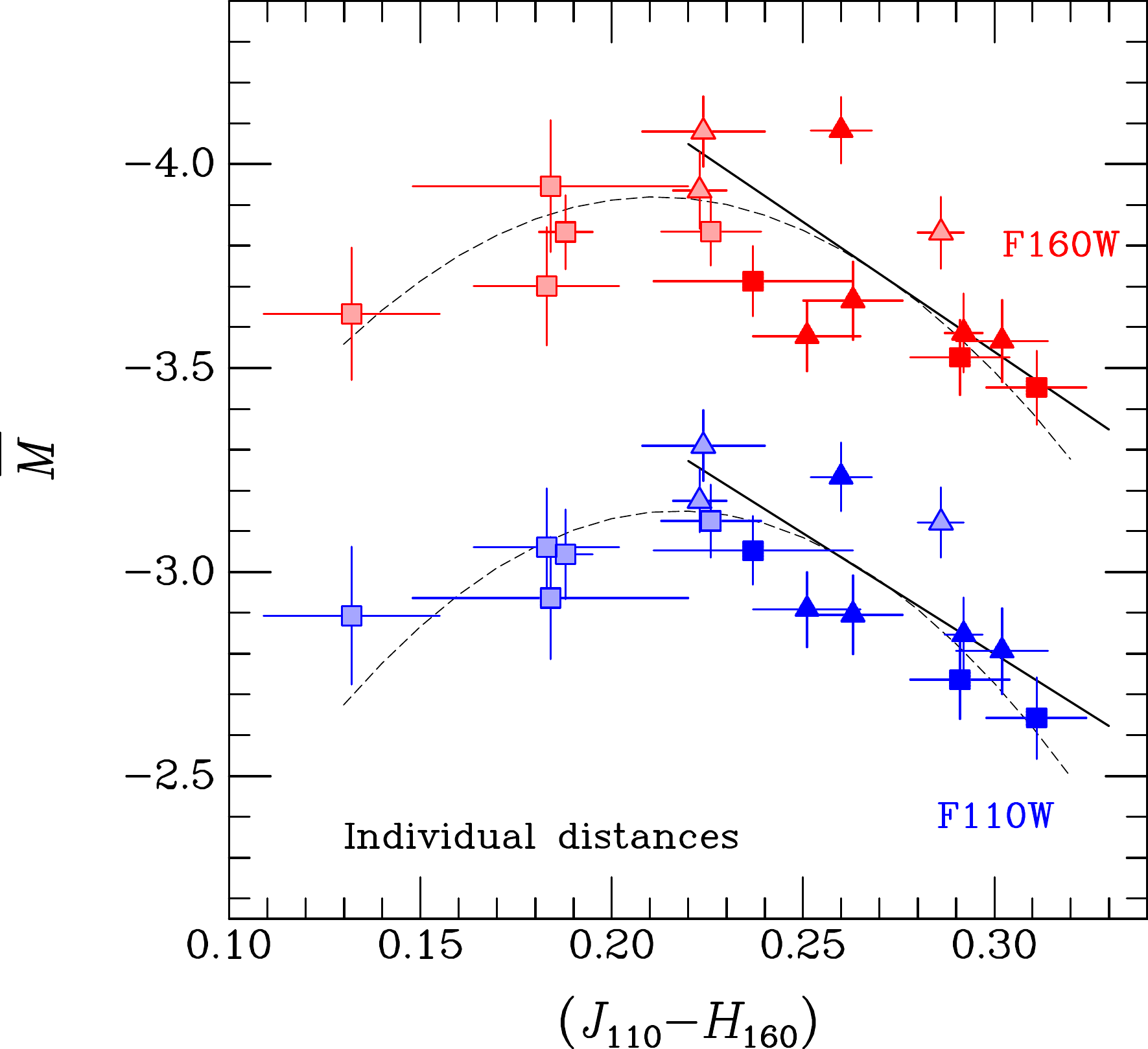}{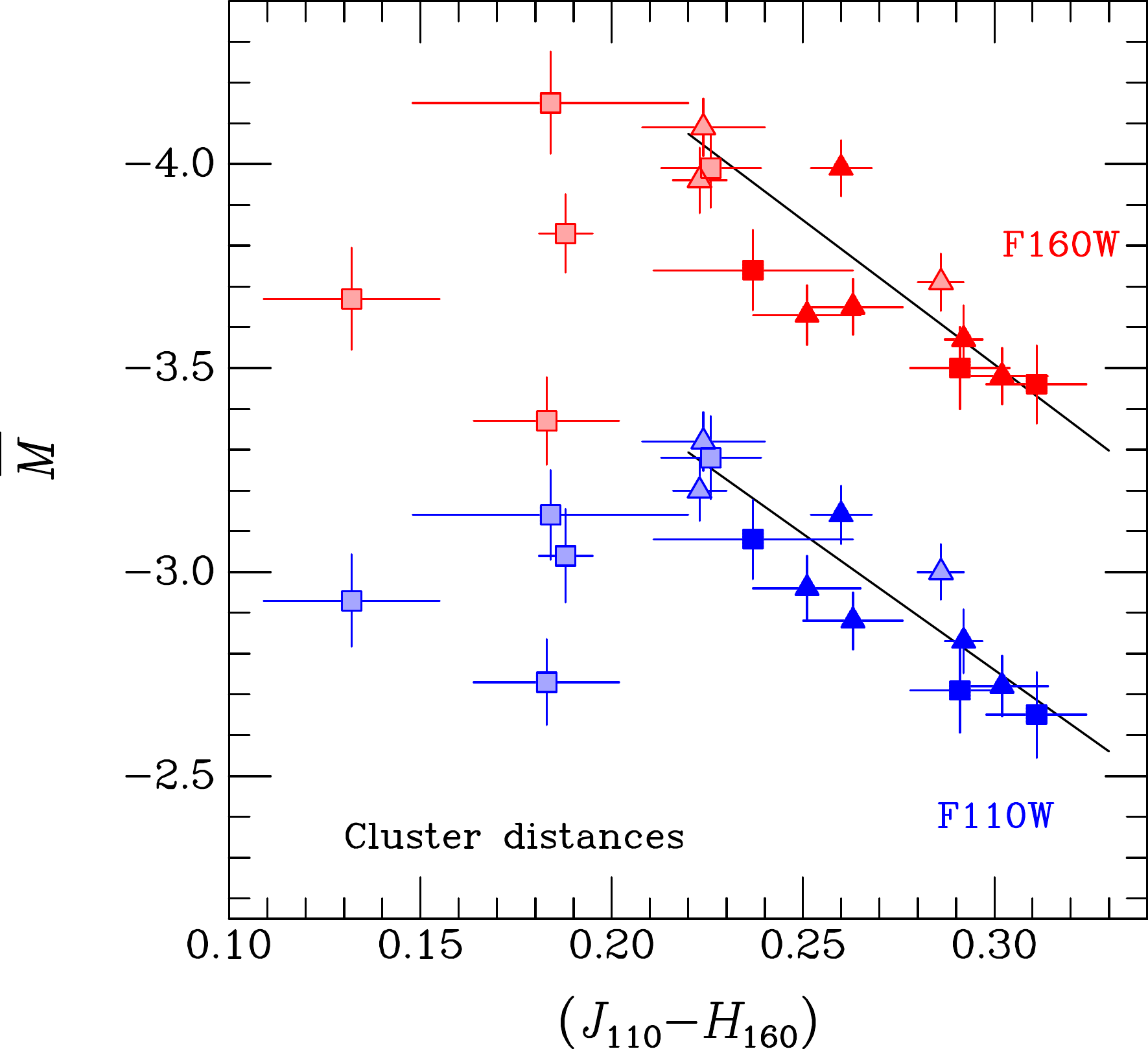}
\caption{Fits to the absolute fluctuation magnitudes \Mbar\ as a 
function of \JH\ color derived using
individual distances (left) and average cluster distances (right). 
Red symbols indicate \Hir\ measurements (top set of points), 
and lower set of blue points are the \Jir\ measurements. Symbol
definitions are the same as in Figure~\ref{calibgz}. \\
\label{calibjh}}
\end{figure*}

As in optical bandpasses,
the intrinsic luminosity of IR fluctuations varies with galaxy color.
The SBF amplitude  is sensitive to the brightness of the most luminous stars in a 
population, and bluer galaxies with a significant component of young 
or intermediate-age stellar populations
have luminous AGB stars that enhance the SBF signal 
(e.g., Jensen et al.\ 2003). An accurate IR SBF distance calibration must 
take into account the brightening of fluctuations at intermediate and 
bluer colors.
As in the previous F160W NICMOS calibration (Jensen et al.\ 2003), 
a linear fit to the red end of the sample was found to best represent the
SBF calibration for distance measurements of typical giant elliptical 
galaxies. 
To determine the best slope of \Mbar\ as a function of galaxy color
(see Figs.~\ref{calibgz} and \ref{calibjh}), 
we adopted an iterative procedure that takes into account the uncertainties 
in both the \Mbar\ and color axes. We started by making an initial
approximate fit ignoring the color uncertainties 
(i.e., using a standard least-squares approach). We computed 
the error ellipse for each point and the distance from the fitted line 
in units of the combined $x$ and $y$ uncertainties.
Because the \zacs\ SBF distances are also a function of \gz, there is
a small correlation between the $x$ and $y$-axis uncertainties when
fitting the individual-distance \Mbar\ values vs. \gz. 
We included the rotation of the error ellipse for that subset of the 
calibration fits.
We then iteratively adjusted the fit coefficients to minimize the 
combined difference between the line and the data points, and then computed
the rms in \Mbar.
This procedure was repeated for each of the filters for both \gz\ and \JH, 
and for the two sets of \Mbar\ values derived from individual and common
cluster distances. 
The coefficients and rms scatter for each fit are listed in 
Tables~\ref{linearcalibrations} and \ref{quadcalibrations}.

The intrinsic scatter in the optical SBF distance measurements is comparable 
to the Virgo and Fornax cluster depths as estimated by 
Blakeslee et al.\ (2009) (0.053 mag for Fornax, 0.085 mag for Virgo). 
Depending upon the location of the galaxy within the cluster, the error in the
estimated distance may be larger when adopting the cluster mean or when using the
individual galaxy distance.
If the typical distance measurement error (including intrinsic scatter about the
stellar population calibration) is less than the magnitude of the scatter from cluster
depth, then it makes sense to calibrate the IR measurements using individual optical
SBF distances.  
If the optical SBF distance measurement errors dominate, then it would be better to
average all the optical SBF measurements and use a common cluster distance
to calibrate the IR measurements. In the case of Fornax, the individual distance
errors are larger than the scatter from cluster depth; the situation is less
clear in Virgo, especially for the bluer galaxies.
Table~\ref{linearcalibrations} presents calibrations computed 
using both individual distance moduli and average cluster distances of 
31.09 mag and 31.51 mag for Virgo and Fornax, respectively 
(Blakeslee et al.\ 2009). 
The linear calibrations using cluster distances are not
significantly different, particularly for the bright ellipticals located near
the centers of the clusters, but the rms scatter is somewhat lower using
the cluster distances. 

The downturn in absolute SBF magnitude at the blue end is dominated by 
the four dwarf galaxies in Virgo. The four bluest galaxies have the
largest color uncertainties and ranges, the lowest SBF $S/N$ ratios,
the lowest galaxy brightness compared to the sky, and the largest radial 
gradients in fluctuation amplitude.
They also show evidence of a wide range in stellar population age and
metallicity, as manifested in their large radial color and 
fluctuation magnitude gradients (the error bars shown in Figures~\ref{calibgz} 
and \ref{calibjh} are larger than the measurement errors, and include the 
range of values due to radial gradients as described below in 
Section~\ref{radialgradients}). These four Virgo galaxies 
were excluded from the linear calibration fits; IC~1919 was excluded
from the \gz\ fits as well.
Because the fluctuation amplitude is significantly lower in
the bluest galaxies in this sample, we also present an alternative
second order polynomial distance calibration fit that can be used for
lower accuracy distance measurements of bluer galaxies
(Figs.~\ref{calibgz} and \ref{calibjh}).
The quadratic fits were computed using the same iterative procedure
that was used to make the linear fits, as described above.
Higher-order fits are not justified given the sample size, measurement
uncertainties, and large population variations between blue dwarf
elliptical galaxies.
The linear calibration is not useful for the bluer galaxies; they 
have too much population variation for SBF to be generally useful as a 
distance indicator. The quadratic calibration may be used to get 
approximate distances for bluer galaxies when necessary.

Calibration coefficients are shown in Tables~\ref{linearcalibrations}
and \ref{quadcalibrations}. 
The coefficients are defined as follows:
\begin{eqnarray}
 \overline{M} \;&=&\; a + b\left[(g{-}z)-1.4\right] + c\left[(g{-}z)-1.4\right]^2 \\
 \overline{M} \;&=&\; a + b\left[(J{-}H)-0.27\right] + c\left[(J{-}H)-0.27\right]^2 
\end{eqnarray}
where $c\,{=}\,0$ for the linear fits.

For $\gz{>}1.2$, researchers measuring distances should use the 
appropriate linear calibrations centered at the mean galaxy color as follows:
\begin{eqnarray}
 \overline{M}_{110} \;&=&\; (-2.946\,{\pm}\,0.015) + (2.16\,{\pm}\,0.15)\left[(g{-}z) - 1.4\right] \label{eq:j_gz} \\
 \overline{M}_{160} \;&=&\; (-3.699\,{\pm}\,0.028) + (2.13\,{\pm}\,0.27)\left[(g{-}z) - 1.4\right] \,. \label{eq:h_gz}
\end{eqnarray}
The quadratic fits may be used for bluer galaxies, bearing in mind that
the intrinsic scatter between blue galaxies is large. Quadratic fits
are not shown for \Mbar\ computed using cluster distances because two 
of the bluest Virgo
galaxies have individual optical SBF distance moduli that differ from 
the mean cluster modulus for Virgo by $-$0.20 and +0.33 mag (IC~3032 and
IC~3025, respectively). 
The differences are 2.4 and 3.9 times larger than the Virgo cluster depth
of 0.085 mag (Blakeslee et al.\ 2009); these galaxies are \textit{probably}
outside the Virgo cluster core and should not be included in a calibration 
based on mean cluster distances.

If \gz\ colors are not available, distances may be computed using
the \JH\ color instead. The scatter in the calibration with \JH\ is
larger than with \gz\ because the color range spanned is 
much smaller. 
For $\JH\,{>}\,0.2$, the following relations centered at the mean 
galaxy color should be used:
\begin{eqnarray}
 \overline{M}_{110} \;&=&\; (-2.964\pm0.032) + (6.7\pm0.9)\left[(J{-}H) - 0.27\right]  \label{eq:j_jh}    \\
 \overline{M}_{160} \;&=&\; (-3.718\pm0.035) + (7.1\pm1.1)\left[(J{-}H) - 0.27\right]  \,. \label{eq:h_jh}
\end{eqnarray}
For magnitudes on the Vega system, subtract 0.7595 mag from \Jir\ AB and 
1.2514 mag from \Hir\ AB for the WFC3/IR filters. The \JH\ color can be 
shifted to the Vega system by adding 0.4919 mag.

The results of our calibration analysis show that IR SBF measurements, especially in
F110W, can produce high-accuracy distance measurements for red early-type galaxies
with $\gz>1.2$ and $\JH>0.2$~mag using Equations~(\ref{eq:j_gz}) through~(\ref{eq:h_jh}).  
It is worth emphasizing again that the linear
calibrations should not be extended to bluer colors: the bluer galaxies, comprising
dwarf ellipticals and low-mass S0s, clearly do not follow extrapolations of
the linear relations in Equations~(\ref{eq:j_gz}) through~(\ref{eq:h_jh}).
Such blue galaxies likely have a wider variety of stellar 
populations than the giant ellipticals and S0 galaxies, and thus
exhibit larger scatter, even with respect to the quadratic fits;
they are not well-suited for highly accurate distance measurements. 
The stellar population implications are discussed in detail below in
Section~\ref{SSPMODELS}.

\begin{deluxetable*}{lcccccc}
\tablecaption{Linear Calibration Coefficients\label{linearcalibrations}}
\tablewidth{0pt}
\tablehead{
\colhead{Calibration} &
\colhead{$a$} &                                              
\colhead{$b$} &                                                 
\colhead{$\chi^2/\rm{dof}$} &
\colhead{rms} &
\colhead{Distances} &
\colhead{$N_{\rm gal}$\tablenotemark{a}}
}
\startdata
\Mbarj\ vs. \gminz & $-2.946\pm0.015$ & $2.16\pm0.15$ &  0.48 & 0.075 & clust & 11\\
\Mbarj\ vs. \gminz & $-2.935\pm0.017$ & $1.90\pm0.16$ &  0.86 & 0.086 & clust & 12\\
\Mbarj\ vs. \gminz & $-2.961\pm0.032$ & $1.86\pm0.32$ &  1.23 & 0.101 & indiv & 11\\
\Mbarj\ vs. \gminz & $-2.951\pm0.034$ & $1.61\pm0.34$ &  1.40 & 0.105 & indiv & 12\\
\\[-5pt]
\Mbarj\ vs. \jminh & $-2.964\pm0.032$  & $6.7\pm0.9$  &  0.94 & 0.092 & clust & 12\\
\Mbarj\ vs. \jminh & $-2.977\pm0.047$  & $5.9\pm1.4$  &  1.76 & 0.130 & indiv & 12\\
\\[-5pt]
\Mbarh\ vs. \gminz & $-3.699\pm0.028$ & $2.13\pm0.27$ &  1.35 & 0.114 & clust & 11\\
\Mbarh\ vs. \gminz & $-3.689\pm0.029$ & $1.90\pm0.28$ &  1.59 & 0.114 & clust & 12\\
\Mbarh\ vs. \gminz & $-3.712\pm0.042$ & $1.76\pm0.42$ &  2.49 & 0.138 & indiv & 11\\
\Mbarh\ vs. \gminz & $-3.704\pm0.043$ & $1.55\pm0.44$ &  2.45 & 0.134 & indiv & 12\\
\\[-5pt]
\Mbarh\ vs. \jminh & $-3.718\pm0.035$  & $7.1\pm1.1$  &  1.35 & 0.121 & clust & 12\\
\Mbarh\ vs. \jminh & $-3.731\pm0.050$  & $6.4\pm1.6$  &  2.44 & 0.162 & indiv & 12
\enddata
\tablenotetext{a}{IC~1919 was excluded from the \gz\ calibration because 
$\gz\,{<}\,1.2$. The \JH\ color for IC~1919 is greater than 0.22, so it was 
included in the \JH\ calibration. We have included \gz\ calibrations with and
without IC~1919 to show the relatively small influence this one galaxy has on
the \gz\ calibration. The recommended calibration in Equations (3) and (4) 
exclude IC~1919.} 
\tablecomments{Linear fits for $\gz\,{>}\,1.2$ or 
$\JH\,{>}\,0.22$. For Vega magnitudes, subtract
0.7595 mag from WFC3/IR \Jir\ and 1.2514 mag from \Hir\ AB. \\}
\end{deluxetable*}

\begin{deluxetable*}{lcccccc}
\tablecaption{Quadratic Calibration Coefficients\label{quadcalibrations}}
\tablewidth{0pt}
\tablehead{
\colhead{Calibration} &
\colhead{$a$} &
\colhead{$b$} &
\colhead{$c$} &
\colhead{$\chi^2/\rm{dof}$} &
\colhead{rms (mag)} &
\colhead{Distances}
}
\startdata
\Mbarj\ vs. \gminz & $-2.991\pm0.004$ & $1.73\pm0.025$ & $4.55\pm0.12$  & 1.11 & 0.115 & indiv \\
\Mbarj\ vs. \jminh & $-2.977\pm0.011$  & $6.3\pm0.4$ & $59\pm6$   & 1.06 & 0.129 & indiv \\
\\[-5pt]
\Mbarh\ vs. \gminz & $-3.723\pm0.003$ & $1.54\pm0.022$ & $3.38\pm0.09$  & 2.17 & 0.128 & indiv \\
\Mbarh\ vs. \jminh & $-3.733\pm0.030$  & $6.9\pm0.9$ & $59\pm15$  & 1.74 & 0.149 & indiv
\enddata
\tablecomments{Quadratic fits for all 16 galaxies. Use only when necessary for 
$\gz\,{<}\,1.2$ or $\JH\,{<}\,0.22$. \\}
\end{deluxetable*}

\subsection{Independent Checks of the IR SBF Calibration\\ Zero Point}

Now, as for several decades, the forefront of progress in the measurement of
extragalactic distances is limited primarily by the uncertainty in the 
calibration zero point (e.g., Freedman \& Madore 2010; Riess et al. 2011). 
We have chosen to calibrate the WFC3 IR SBF
distance scale using the extensive Virgo and Fornax optical SBF measurements
made by Blakeslee et al.\ (2009) and their collaborators. This guarantees that the
IR observational uncertainties and population variations will dominate the
calibration uncertainty, not the precision of the reference distances. 
It does not, however, reduce the systematic zero point uncertainty
present in the optical SBF measurements, which in turn were based on 
\hst\ Cepheid distances (Freedman et al.\ 2001). 
Blakeslee et al.\ (2010) discuss in detail the small offsets between 
several of the largest SBF surveys, including Tonry et al.\ (2001) and
Jensen et al.\ (2003), and the application of metallicity corrections 
to the \hst\ Cepheid distance scale of Freedman et al.\ (2001).
The systematic uncertainty in the SBF distance
scale due to the uncertainty in the Cepheid zero point is about 0.1 mag
(Freedman \& Madore 2010; Blakeslee et al.\ 2010).

One approach for avoiding the Cepheid zero point uncertainty would be to 
use theoretical stellar population model predictions to calibrate the absolute
\Mbar\ in galaxies of varying ages, metallicities, and colors.
This model-based approach would therefore  make SBF a primary 
distance indicator independent of all other distance measurements, dependent
only on our understanding of the luminosities and colors of 
red giants and other evolved stars of a particular age and metallicity. 
Comparisons with several different stellar population models are presented 
below in Section~\ref{SSPMODELS}. 
As will be shown, infrared stellar population models are not sufficiently 
consistent to provide a robust zero point for distance calibration at the
10\% level. At present, we find that observed IR SBF magnitudes are more useful for
constraining stellar population models than the models are for constraining the SBF
distance calibration in the IR.

Another approach is to find other distance indicators that are 
independent of the Cepheid calibration, such as the geometrical distance
to NGC~4258. Water masers orbiting the central black hole in NGC~4258 
have now been used to accurately determine the distance to this galaxy using a 
purely geometrical technique based on the Keplerian orbits of the masers
(Humphreys et al.\ 2013). While SBF magnitudes are best measured in early-type 
galaxies, archival \Hir\ images of the central bulge of NGC~4258 (GO-11570)
provided us with an opportunity to explore the SBF 
calibration independent of the Cepheid distance scale.
Given that an SBF measurement to NGC~4258 could allow us to bypass the 
systematic uncertainty in the Cepheid calibration, we felt it was worth an attempt.
Unfortunately, the presence of clumpy dust and recent star formation prevented
us from achieving an accurate calibration using this galaxy, even when optical
color images were used to identify dusty regions. 
The maser-calibrated fluctuation magnitude is 
significantly brighter (by ${\sim}0.3$ mag) 
than the calibration determined above for the elliptical 
and S0 galaxies in the Fornax and Virgo clusters with similar \gz\ colors (see
Fig.~\ref{calibgz}). 
This result was not a surprise; patchy dust adds to the fluctuation signal,
as does the presence of younger populations containing bright AGB stars. 
Unfortunately, the geometrical distance to NGC~4258 does not provide a 
useful direct calibration of the SBF technique for elliptical galaxies.
Given that optical Cepheid distances to NGC~4258
have now been published by  Macri et al. (2006), Fausnaugh et al.\ (2015),
and Hoffmann \& Macri (2015), 
the future value of NGC~4258 in calibrating IR SBF
is therefore most likely to be through an improved calibration of the Cepheid 
distance scale zero point and metallicity corrections.

Type Ia supernovae are one of the most accurate and widely-used distance
measurement techniques in use today. 
In a recent paper, Cantiello et al.\ (2013) reported WFC3 \Jir\ and \Hir\ 
measurements of the SBF distance to NGC~1316, a type Ia supernova host 
galaxy in the Fornax cluster. 
They used the Jensen et al.\ (2003)
F160W SBF calibration for NICMOS (including a metallicity correction to the
Freedman et al.\ 2001 Cepheid zero point) and
applied a 0.2 mag offset to the NICMOS zero point to account for the
difference in filter width between NICMOS and WFC3/IR, based on predictions
of the SPoT stellar population models (Raimondo 2009; Raimondo et al.\ 2005).
To avoid uncertainties arising from  differences between the methods 
used by Cantiello et al.\ and those used herein, we repeated the SBF analysis 
for NGC~1316 using the original WFC3/IR data (GO-11691) and the procedures 
described above. 
Our measured SBF magnitudes are listed in Tables~\ref{jmeasurements} and 
\ref{hmeasurements}; they
are consistent with the Cantiello et al.\ (2013) values within the stated 
uncertainties. 
This galaxy is not an ideal SBF candidate due to the 
presence of extensive patchy dust near the center, but the regions farther
out from the center appear relatively clean and the SBF signal is very strong. 
If we adopt the Blakeslee et al.\ (2009) $z$-band SBF distance (Table~\ref{absmags}),
the \Mbar\ values we find for NGC~1316 are brighter than the
calibration prediction by ${\sim}$0.3 to 0.5 mag.
The Tonry et al.\ (2001) ground-based
$I$-band SBF measurement of $31.66\pm0.17$ supports the 
Blakeslee et al.\ (2009) distance for NGC~1316. 
If NGC~1316 is located at the same distance as the core ellipticals in Fornax,
then it appears to have a significant population of younger  
AGB stars, most likely the result of star formation that took place during
a major merging episode a few Gyr ago, that biases the IR SBF magnitude. It
may also have additional undetected dust contributing to the fluctuations.
On the other hand, the IR and optical SBF distances are inconsistent
with the published type Ia supernova distances, which place NGC~1316 ${\sim}0.25$ mag
closer than the Fornax cluster core. 
If we adopt the Ia supernova distance modulus for the three 
normal supernovae published by Stritzinger et al. (2010) of 
$31.248\,{\pm}\,0.034\,{\pm}\,0.04$ instead of the $z$-band SBF distance, 
our IR SBF measurements would be much more consistent with the 
elliptical galaxy calibration. 
Additional work is needed to reconcile the SBF and 
supernova distances to this galaxy.

The IR SBF measurements of the maser host galaxy NGC~4258 and
the supernova host galaxy NGC~1316 are not sufficiently accurate at present to reduce
the systematic uncertainty in the IR SBF distance scale calibration to less than
the 10\% that we currently inherit from the Cepheid calibration of the optical
SBF distances used for this study (Blakeslee et al.\ 2009, 2010). These galaxies
have significant patchy dust and recent star formation, both of which enhance
the IR SBF signal over what is typically observed in quiescent elliptical galaxies.

\subsection{Comparison of WFC3 and NICMOS SBF Magnitudes}

\begin{figure*}
\epsscale{0.85}
\plottwo{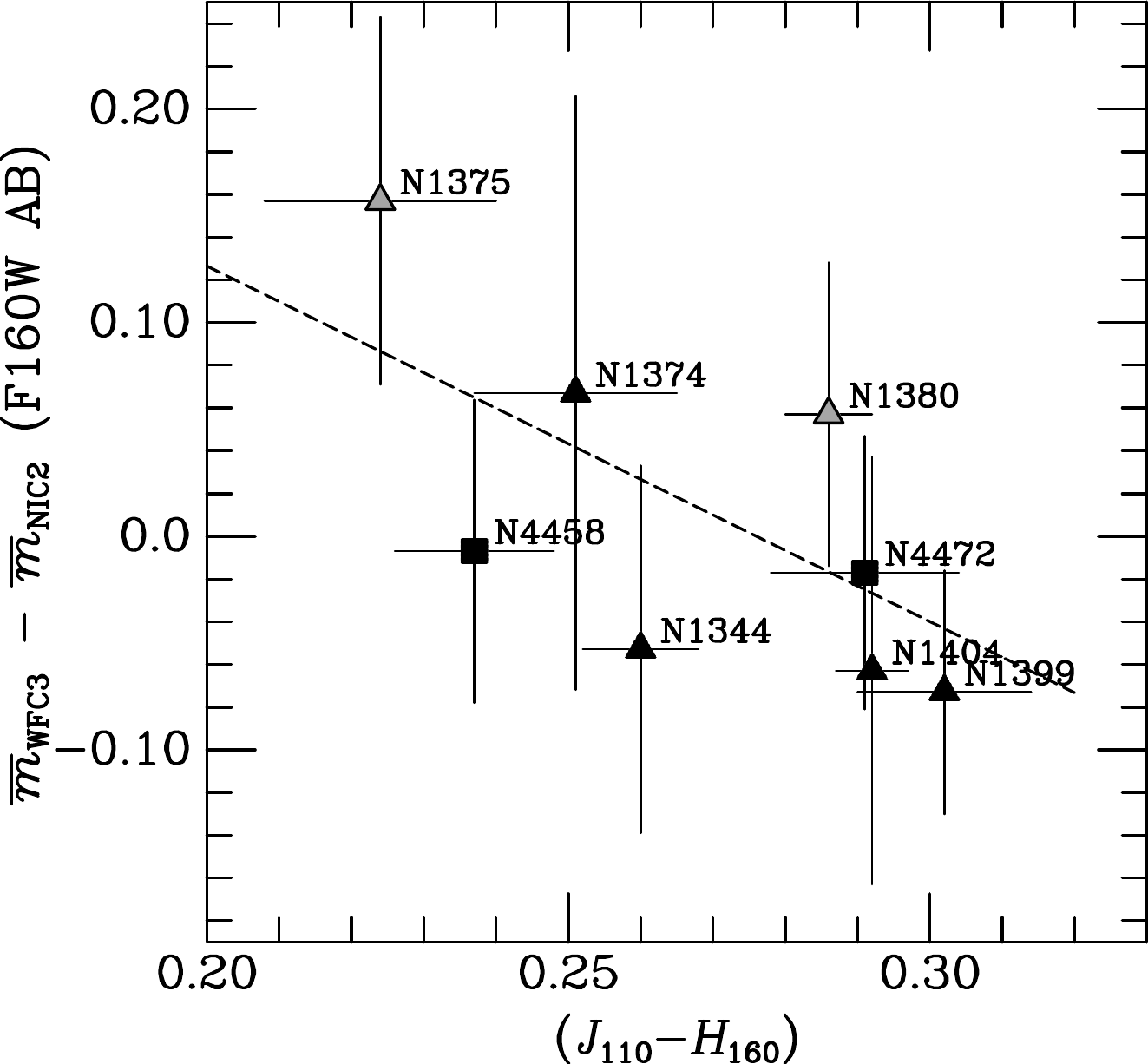}{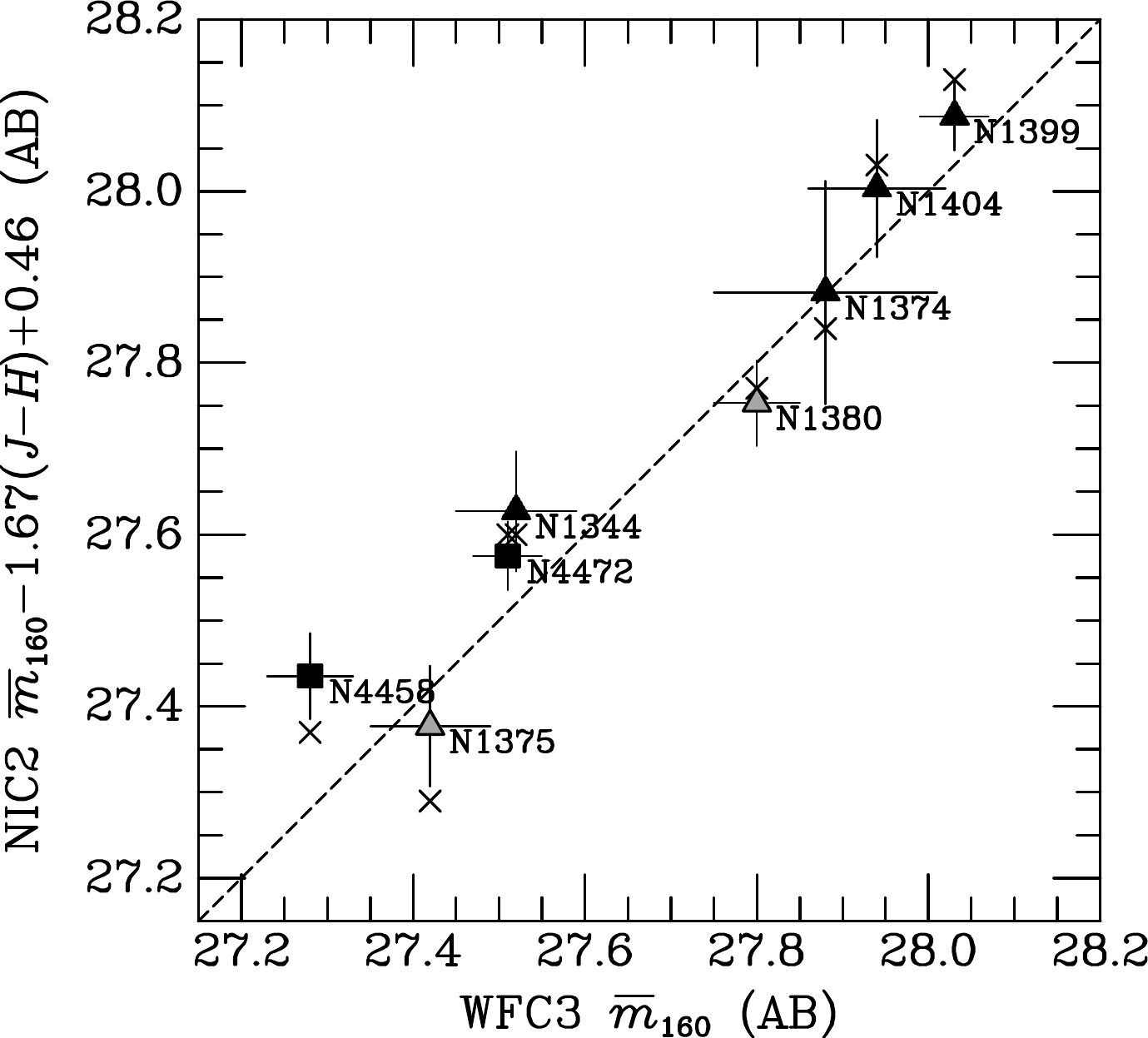}
\caption{Comparison of NICMOS NIC2 and WFC3/IR \Hir\ SBF measurements. 
The left panel shows the fit to the color term used to correct the points 
in the right panel, which shows the corrected NIC2 SBF magnitudes plotted as a
function of the WFC3/IR SBF magnitudes. 
For reference, the SBF magnitudes with no color correction 
are shown with $\times$ symbols.
\\
\label{niccomparison}}
\end{figure*}

Eight galaxies in our current sample, mostly in the Fornax cluster, 
were included in the NICMOS NIC2 calibration of \Mbarh\ (Jensen et al.\ 2003). 
A comparison of the SBF magnitudes is shown in Figure~\ref{niccomparison}.
To make the comparison using AB magnitudes, we added 1.313 mag to the 
published NIC2 SBF magnitudes to shift from the Vega magnitude system
used by Jensen et al.\ (2003) to the AB mag system.\footnote{http://www.stsci.edu/hst/nicmos/documents/handbooks/DataHandbookv8/ nic\_ch5.9.3.html}

The difference between NIC2 and WFC3/IR F160W SBF magnitudes shows a modest
color dependence (left panel in Fig.~\ref{niccomparison}). We
fitted the slope (the color term in the conversion of NIC2 SBF magnitudes to
WFC3/IR) including uncertainties in both \JH\ color and \mbarh\ SBF
magnitudes; the rms for the fit is 0.055 mag and the $\chi^2$ per degree 
of freedom is 0.66. 
The right panel of Figure~\ref{niccomparison} compares the 
NIC2 F160W apparent SBF magnitudes to WFC3/IR, with and without the color 
correction. 
The dashed line in the right panel of Figure~\ref{niccomparison} is not a 
fit, but shows the 45-degree perfect correlation line. 
SBF \mbarh\ measurements made using NIC2 prior to the installation of
the NICMOS cryocoolers may be compared to WFC3/IR measurements using the
relation
\begin{equation}
 \overline{m}_{160} \;=\; \overline{m}_{\rm 160,NIC2} - 1.67\JH + 0.46
\end{equation}
(all AB magnitudes). Because we are comparing apparent fluctuation
magnitudes directly, we do not need to be concerned about differences in
distance scale calibrations  between the two studies or cameras.

\section{Stellar Population Models\label{sspmodels}}

We turn now to
how IR SBF measurements, with their sensitivity to red giant branch and
intermediate-age AGB stars, can expose interesting differences between 
galaxies with different star formation histories and better constrain
single-burst stellar population models.

The brightening of the SBF magnitudes in elliptical and S0 galaxies
at intermediate colors seen in the centers of Figures~\ref{calibgz} 
and \ref{calibjh}, 
and the subsequent drop in SBF brightness in the bluest dwarf ellipticals
in this sample, provide powerful new constraints for stellar population models,
which have only been compared to redder galaxies in previous near-IR SBF studies.
The scatter among the bluest galaxies in our sample is much larger than the 
observational uncertainties, and many of these galaxies exhibit significant 
radial gradients in IR fluctuation magnitude, color, or both.
The current sample includes bluer and fainter galaxies than are 
typically targeted for SBF distance measurements, and the breaking of the 
age-metallicity degeneracy in near-IR fluctuations provides a unique
opportunity to explore the stellar populations in these galaxies.

Single-burst stellar population (often abbreviated SSP) models with 
constant age and metallicity are frequently used to interpret broad-band 
colors and other properties of unresolved stellar populations in distant 
galaxies, including SBF magnitudes. 
These models are usually calculated by integrating collections of 
properly-weighted isochrones, and can be used to compute predicted SBF 
magnitudes directly without any need to link the apparent SBF magnitudes to an
external distance scale calibration. Examples of theoretical SBF comparisons
include Worthey (1993, 1994), Liu et al.\ (2000, 2002), 
Blakeslee et al.\ (2001), Cantiello et al.\ (2003), Raimondo et al.\ (2005), 
Mar\'in-Franch \& Aparicio (2006), Biscardi et al.\ (2008), and 
Lee et al.\ (2010).

We compared our SBF measurements to three recent sets of SSP 
models for which \Jir\ and \Hir\ SBF magnitudes have been computed.
The purpose of these comparisons is to explore the limitations of our
SBF distance calibration and provide input to researchers working 
to improve stellar population models, particularly for understanding
galaxy evolution, when the observations may not so easily distinguish the
effects of age and metallicity as our near-IR SBF measurements do. 
While we compared our IR SBF measurements to single-burst population models, 
real galaxies are composed of composite stellar populations with potentially many 
bursts of star formation. Because the fluctuations are dominated by the most 
luminous stars weighted as $L^2$, they are even more strongly weighted towards 
young, luminous populations than are broad-band galaxy colors 
(Tonry \& Schneider 1988). 
A composite population model will therefore predict an SBF magnitude 
close to that of the younger (or brighter) model component, even when the young population 
is only a small fraction (${\sim}10$\% to 20\%) of the galaxy by mass 
(Jensen et al.\ 2003; Blakeslee et al.\ 2001; Liu et al.\ 2002). 
The comparison to SSP model ages shown in this section should be considered the time 
since the most recent episode of star formation, not the average age of the dominant
stellar population by mass.

\subsection{Teramo BaSTI Models}
The first set of stellar population models we consider here are based on the 
Teramo BaSTI models\footnote{\url{http://193.204.1.62/index.html}} 
using a standard Salpeter initial mass function (IMF) with a low-mass 
cutoff of 
0.5~$M_{\odot}$ (Lee, Worthey, \& Blakeslee\ 2010; Pietrinferni et al.\ 2004, 
2006; Cordier et al.\ 2007). 
We compared our observed fluctuation
magnitudes \Mbarj\ to absolute fluctuation magnitudes computed
for two variants of the BaSTI models: the solar-scaled abundance ratio models
without convective overshoot, and the $\alpha$-enhanced version of the
models described by Lee et al.\ (2010).  The latter models have a mean
[$\alpha$/Fe] of $\sim\,$0.4~dex, but with physically-motivated variations
in the abundance ratios of the individual $\alpha$ elements 
(i.e., O, Mg, Si, S, Ca, Ti); the [Fe/H] abundances in these models have been 
correspondingly reduced to keep a fixed [Z/H].

\begin{figure}
\plotone{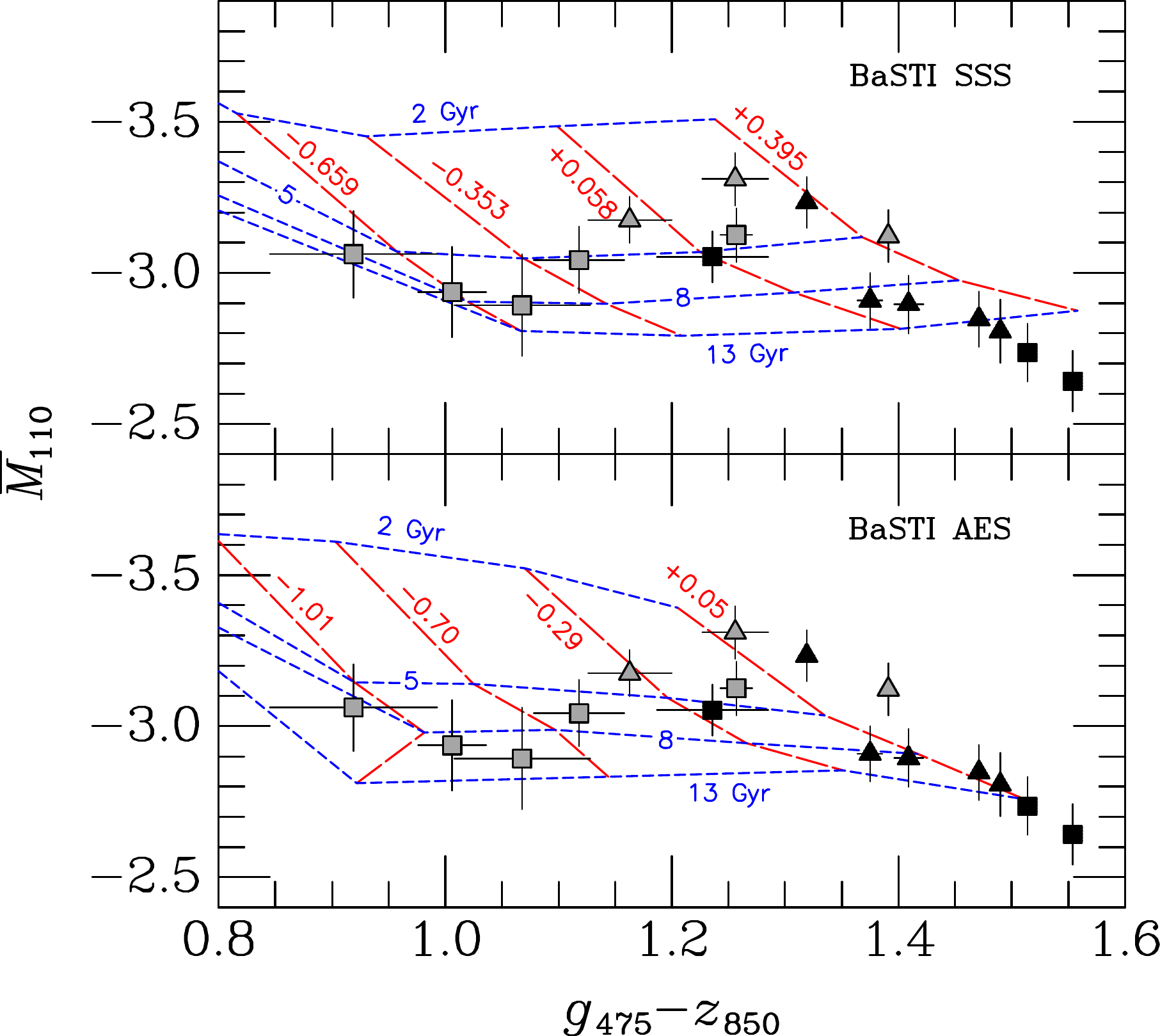}
\caption{Teramo BaSTI models compared to \Mbarj\ as a function of
galaxy \gz\ color for two metallicity variants of the models computed
using the individual galaxy distances: 
the ``SSS'' models (top panel) are solar-scaled
abundances spanning the range from [Fe/H]$\,{=}\,-0.659$ to +0.395.
The ``AES'' models (lower panel) have enhanced $\alpha$ element compositions 
and reduced [Fe/H] to make the overall metallicity $Z$ the same as the 
solar-scaled models shown in the top panel. 
The range of [Fe/H] spanned by the AES models is $-$1.01 to +0.05. 
Lines of constant metallicity are shown with red dashed lines, and lines of 
constant age from 2 to 13 Gyr are plotted with dotted blue lines.
\label{teramo110}}
\end{figure}

The solar-scaled models shown
in the top panel of Figure~\ref{teramo110} do not match the reddest giant
ellipticals as well as the $\alpha$-enhanced models shown in the lower panel
(the solar-scaled models are included here to provide a point of reference with 
past SBF-model comparisons that only used solar-scaled metallicity models). 
Both model variations suggest that the intermediate-color galaxies in our
sample have younger populations, as expected due to the presence of 
intermediate-age AGB stars. 
The models imply that the bluest galaxies are old and metal-poor. 
While we have chosen to show the comparison for \Mbarj\ and
\gz, the conclusions are the same when we compare the models to \Mbarh\ 
instead of \Mbarj\ or \JH\ instead of \gz.

\subsection{Teramo SPoT Models}

The second set of models was developed by the Teramo SPoT group
(version BaSeL3.1, Raimondo 2009; Raimondo et al.\ 2005).
The Teramo SPoT models pay special attention to the thermally-pulsating
asymptotic giant branch (TP-AGB) stars, particularly in young to 
intermediate-age populations. The SPoT model SBF magnitude predictions
in the near-IR have been empirically compared to a variety of clusters in the
Large Magellanic Cloud and have been shown to match empirical measurements
of SBF magnitudes, integrated magnitudes, star counts, and colors 
(Raimondo 2009; Cantiello et al.\ 2007; Cantiello 2012).

\begin{figure}
\plotone{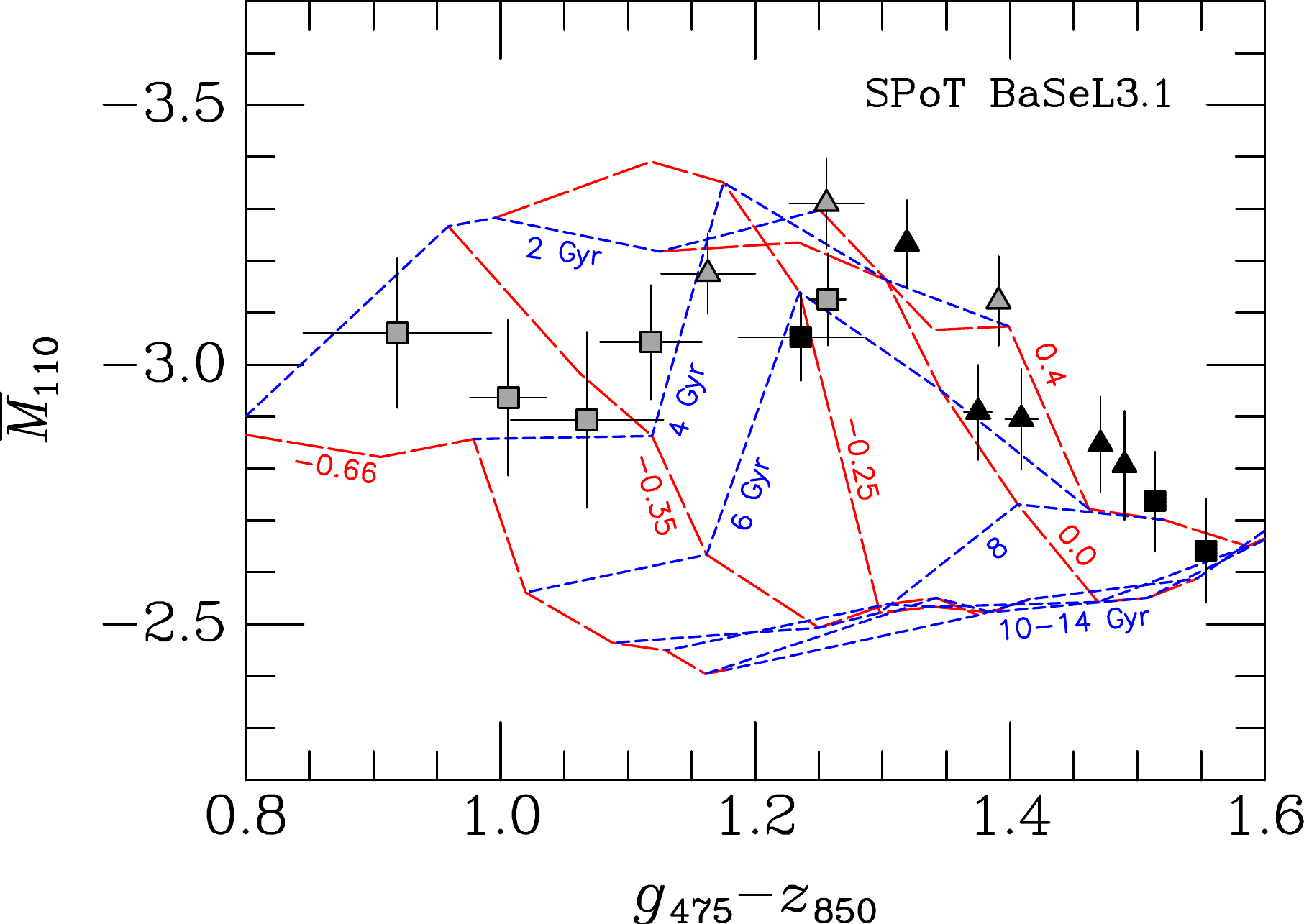}
\caption{Recent versions of the Teramo SPoT models compared to \Mbarj\ measurements 
as in Figure~\ref{teramo110}. The metallicity range for these models spans
[Fe/H]$\,{=}\,-0.66$ to +0.4 and the ages from 2 to 14 Gyr.
The lines and symbols are defined in the
same way as in Figures~\ref{calibgz} and \ref{teramo110}.
\label{spot110}}
\end{figure}

The SPoT models in Figure~\ref{spot110}\ differ significantly from the BaSTI models
in Figure~\ref{teramo110}, although they are based on the same 
stellar evolutionary tracks as the BaSTI models in the upper panel of Fig.~\ref{teramo110}.
The SPoT models derive SBF magnitudes in a procedure that allows modelers to
statistically combine various stellar population models produced by stochastic
variations in the number and properties of bright and rare stars, including the
TP-AGB and horizontal branch populations in intermediate-age and old stellar
populations.
Overall, the SPoT and BaSTI models agree for the red and intermediate-color
ellipticals, with somewhat fainter SBF magnitudes for the oldest and most metal-rich
giant ellipticals.
The SPoT models predict a larger spread in SBF magnitude at younger ages and imply
that the bluer galaxies in the sample are all younger than about 5~Gyr, in contrast with the
BaSTI models, which span a narrower range in \Mbarj\ and imply ages greater than
5~Gyr for the bluest galaxies.

\subsection{Padova Models}

\begin{figure}
\plotone{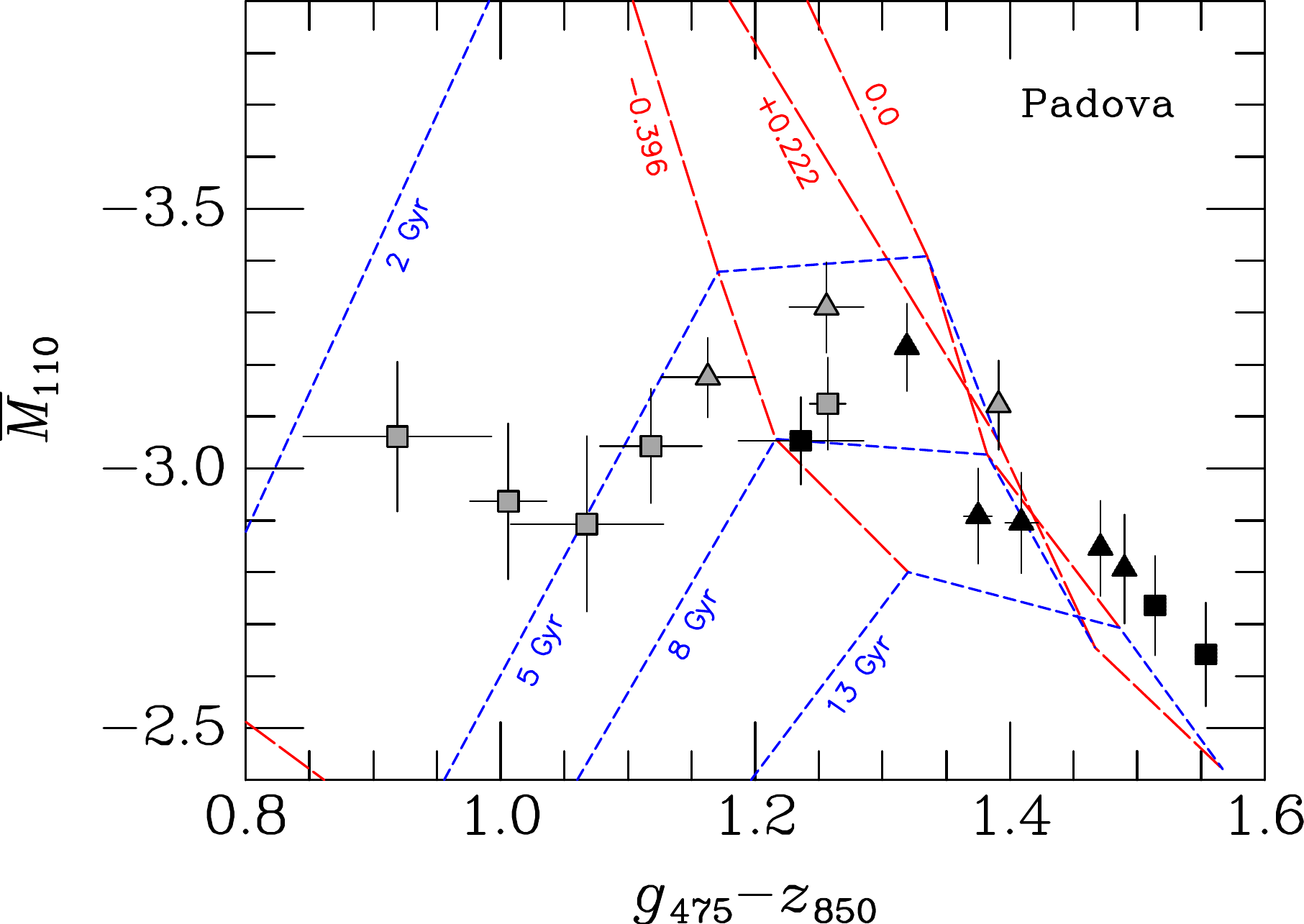}
\caption{Stellar population models based on the Padova isochrones
compared to \Mbarj\ measurements computed using individual SBF distances.
The symbol definitions are the same as in Figure~\ref{calibgz}.
Lines of constant metallicity from [Fe/H]$\,{=}\,-0.705$ (in the extreme lower
left corner) to 0.0 and +0.222 (overlapping at right) are shown
with red dashed lines. Lines of constant age from 2 to 13 Gyr are plotted
with dotted blue lines. 
\\
\label{padova110}}
\end{figure}

The third set of models are based on the Padova isochrones 
(Fig.~\ref{padova110}), which include
sophisticated handling of the TP-AGB evolutionary phase. The Padova
tracks use solar-scaled metallicity abundance ratios and do not
include $\alpha$-element enhancement. SBF magnitudes were computed by 
Lee et al.\ (2010) using the evolutionary tracks of Marigo et al. (2008). 

The Padova SSP models show the largest spread in \Mbarj\ of all the models 
considered here. At the red end, the giant ellipticals again agree 
with old, metal-rich population models, with the intermediate bluer galaxies 
having brighter fluctuations and younger populations. 
The bluest galaxies in the sample
are consistent with ages intermediate between the BaSTI and SPoT models,
in the range 3 to 7 Gyr. Because of their larger spread in SBF magnitude predictions,
the Padova models, as compared to the other sets of models,
 suggest that the data provide better discrimination
between ages and metallicities.

\subsection{Fluctuation Colors}

\begin{figure}
\plotone{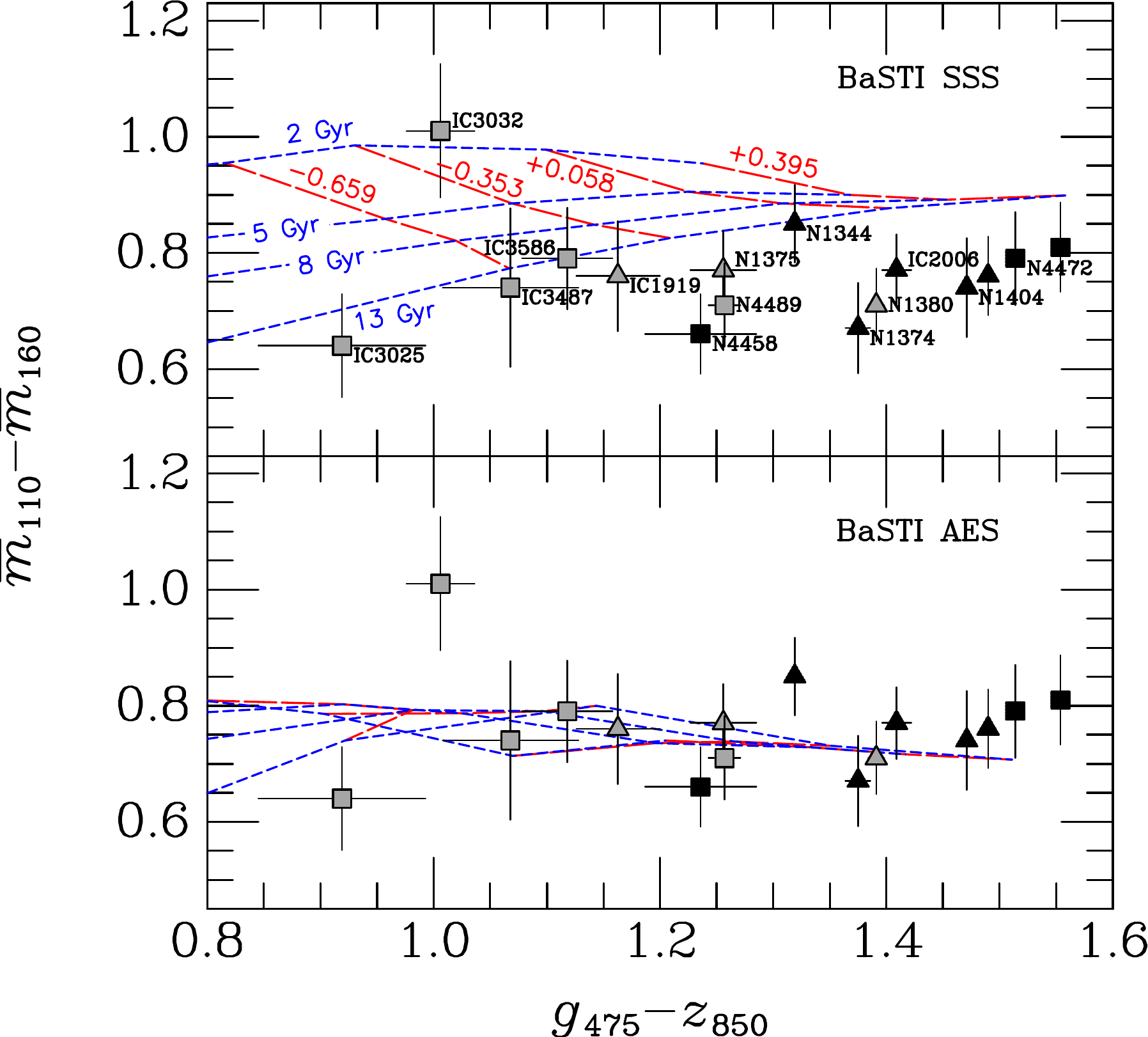}
\caption{Distance-independent IR fluctuation colors for the solar-scaled and 
$\alpha$-enhanced Teramo models. Blue dotted lines indicate lines of constant
age, from 2 to 13 Gyr. Red dashed lines show constant metallicity tracks,
from [Fe/H]$\,{=}\,-0.659$ at the left to +0.395 at right (upper panel)
and [Fe/H]$\,{=}\,-1.01$ to +0.05 (lower panel). 
The two points that are not labeled for 
clarity are NGC~1399 in Fornax and NGC~4636 in Virgo. 
Symbol definitions are the same as in Figures~\ref{calibgz} and \ref{teramo110}.\\
\label{teramo110-160models}}
\end{figure}

SBF measurements at two wavelengths can be used to eliminate 
uncertainties resulting from distance error; for instance, the ``fluctuation color''
\colorbar\ can be compared to stellar population models in a distance-independent way. 
Figure~\ref{teramo110-160models} compares \colorbar\ to the Teramo BaSTI
solar-scaled and $\alpha$-enhanced models as a function of \gz.
As discussed by Lee et al.\ (2010), the predicted SBF magnitudes are sensitive
to $\alpha$-element abundance mostly because of the effects of oxygen-enhancement on the
upper red giant branch and AGB phase.
The data in Figure~\ref{teramo110-160models} agree on average with the
$\alpha$-enhanced models relatively well, but only poorly with the
solar-scaled models.  However, this set of $\alpha$-enhanced 
models also predicts a narrower range of \colorbar\ than is observed.
In particular, the low-mass dwarf IC\,3032 in Virgo agrees better with the locus of
the solar-scaled models; this may indicate real variation in $\alpha$-element
abundance ratios among the sample galaxies.
We conclude that fluctuation colors
can provide useful information on elemental abundance ratio trends with age and
metallicity in elliptical galaxies, independent of the uncertainties in the distance
calibration.
In addition, comparisons with other observables that are sensitive to age,
metallicity, or $\alpha$-enhancement should provide powerful joint constraints for 
future stellar population models.

\subsection{Radial SBF Gradients\label{radialgradients}}
\begin{figure}
\plotone{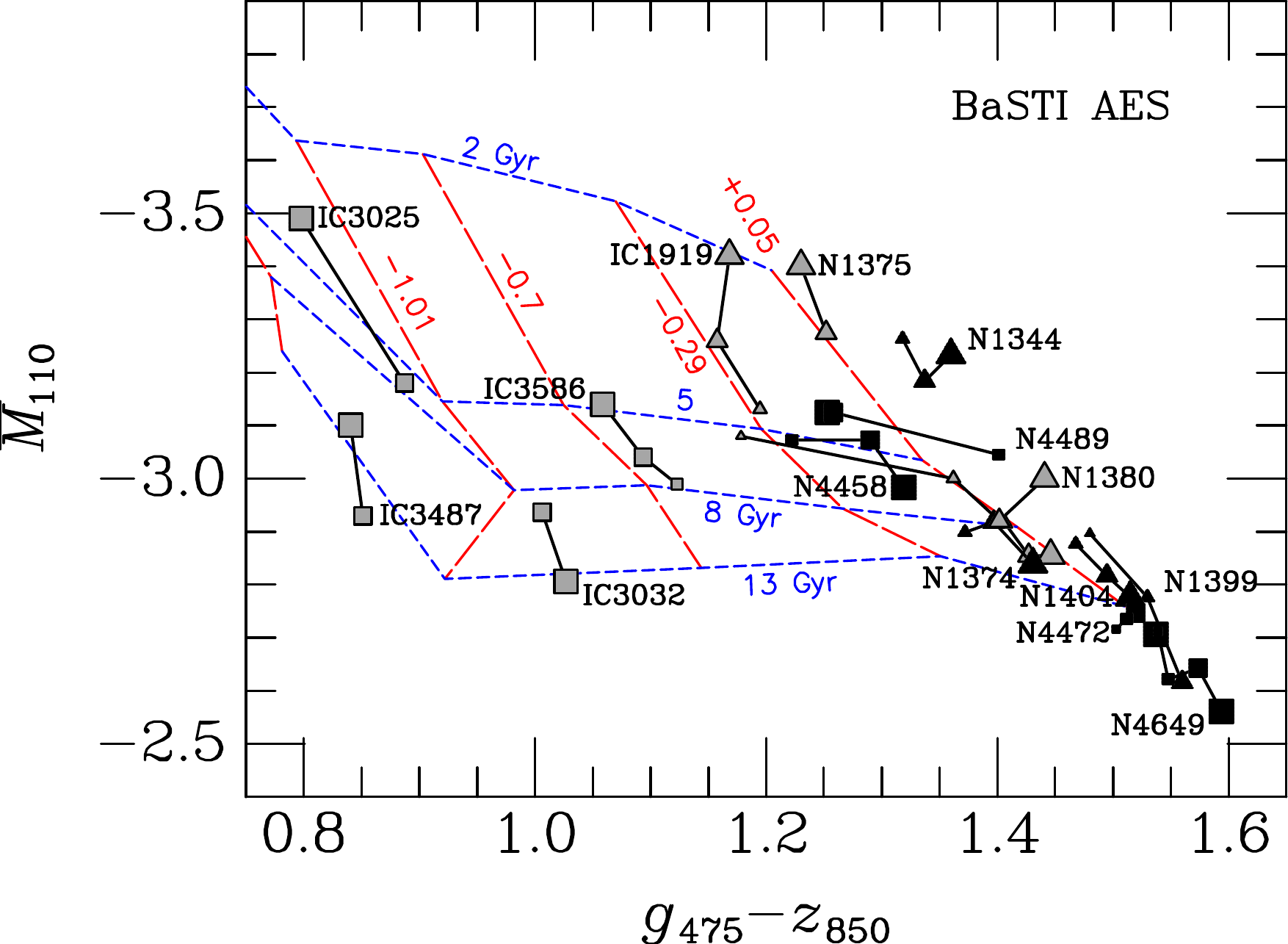}
\caption{Radial SBF and color gradients for the survey galaxies, with larger
symbols corresponding to the inner annuli in which SBFs were measured and
smaller symbols being outer regions. Black lines link measurements in
the same galaxy.
The measurements are compared to the BaSTI $\alpha$-enhanced models;
the conclusions are the same for the other models because they all 
have lines of constant metallicity with similar slopes for this color range.
IC~2006 is not labelled but falls very close to NGC~1374.
\label{teramo110-radial}}
\end{figure}

To further explore the origins of the scatter in SBF magnitude among the 
low-luminosity blue dwarf ellipticals, 
we measured the radial behavior of the SBF amplitude.
Many of the galaxies in our sample, particularly the low-luminosity galaxies,
are quite elliptical. To get the cleanest gradient measurements possible,
we repeated the SBF analysis and measured \gz\ colors in elliptical 
annuli for a subset of the galaxies with significant \Mbar\ gradients
(IC~1919, NGC~1375, NGC~1380, IC~3025, IC~3487, and IC~3586; 
see Fig.~\ref{imagefigure}). For the rest of the sample we used the circular
annulus SBF and \gz\ color measurements previously used for the calibration.
The results are plotted in Figure~\ref{teramo110-radial}.

The comparison of radial SBF gradients to the models shows a significant
difference between the giant ellipticals and the smaller galaxies. 
Most of the galaxies show a gradient sloping from upper left to
lower right in Figure~\ref{teramo110-radial}, roughly along lines of constant
metallicity. While the various models look quite different in detail,
they all have lines of constant metallicity sloping in approximately the 
same direction at the relevant colors. 
The lower-luminosity galaxies have centers (larger symbols in
Fig.~\ref{teramo110-radial}) consistent with younger stellar populations 
(brighter fluctuations) and nearly constant metallicities: 
IC~1919, IC~3025, IC~3487, IC~3586, and NGC~1375 all have brighter
fluctuations near their centers. 
In contrast, the giant ellipticals on the
red end of the diagram tend to show older, and sometimes more metal-rich,
populations in their centers: IC~3032, IC~2006, NGC~1374, NGC~1399, 
NGC~1404, NGC~4458, and NGC~4649 have fainter fluctuations towards their
centers. 
Four of the sixteen galaxies (NGC~1380, NGC~1344, NGC~4458, and NGC~4472) 
appear to have color and SBF gradients consistent with little or no age
variation. 
The majority of the galaxies
show gradients consistent with primarily age variations, however.

The IR SBF and color gradients hint at differing formation histories for the 
galaxies.
As noted above, low-luminosity blue elliptical galaxies usually have older
populations at large radii, and thus to have formed stars more recently near their 
centers from metal-poor gas, while the giant ellipticals formed stars in their 
cores long ago from enriched gas. 
Optical studies of SBF gradients, in contrast, suggest that the  outer parts of 
giant ellipticals
have colors and fluctuation magnitudes consistent with lower metallicity
populations and little or no age gradient (Cantiello et al.\ 2005, 2007).
Our IR measurements do not cover as large a range in radius as the optical
studies because of the smaller field of view spanned by the IR detectors and
the brighter sky background.
Because the
data here are confined within the effective radius of the several largest galaxies
(Table~\ref{tab:props}), we cannot derive strong constraints on 
the large-scale stellar population gradients in these giant galaxies. 
Interestingly, the physical size ($r\lesssim\,$5~kpc) of 
the radial region probed by our SBF data is similar to
that of the luminous dense cores observed at high redshift (e.g., van Dokkum et
al.\ 2010; van der Wel et al.\ 2014), which are believed to be the seeds
around which massive modern-day ellipticals were assembled through hierarchical
merging and accretion.  Our results on the old ages and relative homogeneity of the
stellar populations of the giant ellipticals within the central region are thus in
line with this scenario for
early-type galaxy evolution, if the elliptical galaxies in our sample may be
compared with the high-redshift counterparts (Fritz et al.\ 2014).  
At much larger
radii in giant ellipticals, evidence exists from optical and IR color gradients
that the stellar populations at several effective radii are older than the mean
stellar age in the core (La Barbera et al.\ 2012; but see also Greene et al.\ 2015).
While our data do not probe such
large radii in the massive galaxies, we find the same trend for the lower mass
galaxies where our measurements do reach beyond the effective radius.  It would be
interesting to extend the IR SBF gradients to correspondingly large radii in the
massive ellipticals to see if these also show evidence for larger ages in 
the galaxy outskirts.

Crucially for the purpose of distance measurement, the radial and population variations in the
reddest galaxies cause SBF magnitude variations \emph{along} the direction 
of the linear calibration. These galaxies
are ideally suited for distance measurement because variations from galaxy
to galaxy in age or metallicity are adequately accommodated by the calibration
slope with \gz. 
On the other hand, the radial gradients in SBF magnitude in the bluer dwarf galaxies 
(the gray symbols in the left half of Fig.~\ref{teramo110-radial} 
are perpendicular to the quadratic calibration relation, leading to \emph{greater} scatter 
in the distance calibration. 
These low-luminosity dwarfs show a wide range of ages and metallicities, 
with several showing evidence of recent star formation near their centers, as
revealed by their SBF magnitudes.

Only one set of SBF measurements and models is shown in 
Figure~\ref{teramo110-radial}; the general conclusions, however, are consistent for
all the models discussed herein. The radial variations are consistently sloped
along lines of constant metallicity for all the models, so even though the
models might not agree on the specific age and metallicity of a particular
galaxy, the trend towards younger ages in the centers of bluer dwarf galaxies
is consistent for all the models.

\subsection{Line Index Age and Metallicity Constraints}

\begin{figure}
\plotone{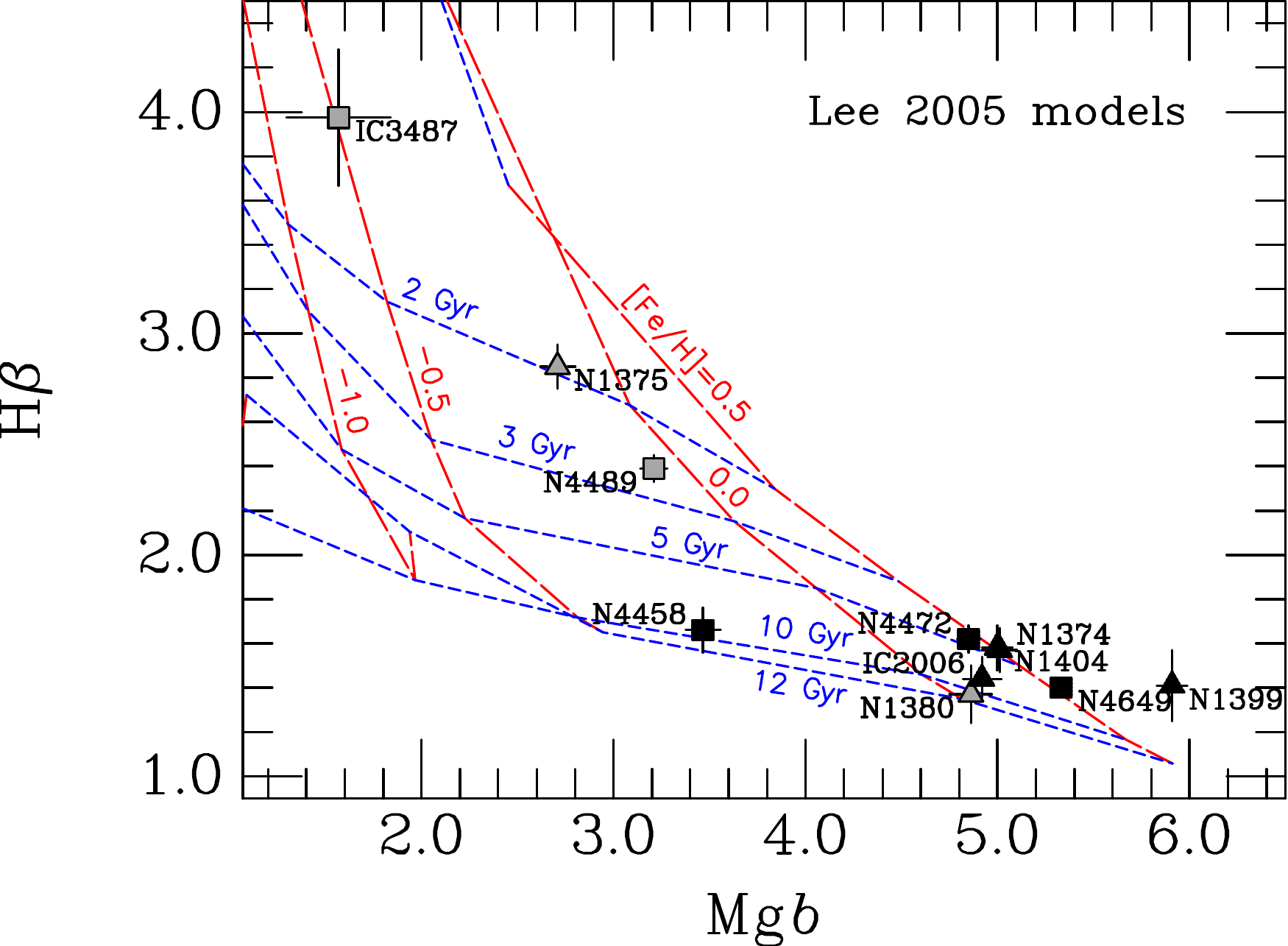}
\caption{Lick index measurements for eleven of the galaxies in our sample
(Kuntschner 2000; Trager et al.\ 2000; Caldwell et al.\ 2003) as compared
with the $\alpha$-enhanced models from Lee \& Worthey (2005). 
H$\beta$ is more age-sensitive, while Mg$b$ is metallicity-sensitive. Such
data provide an independent way of checking the SBF model comparison
conclusions; see text.\\ 
\label{lickindexfigure}}
\end{figure}

Eleven of the galaxies in our sample have published $H\beta$ and Mg$b$ Lick
index measurements from Kuntschner (2000), Trager et al.\ (2000), and Caldwell et
al.\ (2003). The former index is more sensitive to age, and the latter to metallicity.
Figure~\ref{lickindexfigure} compares these data with the $+$0.3~dex $\alpha$-enhanced 
model predictions from  Lee \& Worthey (2005).
The line index measurements generally sample different regions of the galaxies than
our SBF data, but we see again that the reddest ellipticals 
agree with the old, metal-rich population models. 
The youngest galaxies as determined using absorption lines do not always
agree with the SBF models. IC~3487 has the strongest H$\beta$ index but the
faintest \Mbarj\ among the bluer galaxies, implying a relatively older age
compared to NGC~1375, which has somewhat weaker H$\beta$ and brighter \Mbarj\
(although there is considerable variation between SBF models at the 
youngest ages). Further work is needed to explore the radial behavior
of Mg$b$ and H$\beta$ as compared to our SBF measurements on the same scales,
particularly since we detect a significant
radial age gradient in many of the lower-luminosity galaxies.

\subsection{Model Comparison Summary}
The SBF models shown in Figures~\ref{teramo110}, \ref{spot110}, and
\ref{padova110} agree for old, metal-rich populations such as those
commonly found in giant elliptical galaxies, and for which extensive
comparisons have been made in the past (e.g., Jensen et al.\ 2003; 
Cantiello et al.\ 2012; Fritz 2012).
At younger ages and lower metallicities, the three sets of models are
significantly different, and the conclusions we draw about the ages of the blue
dwarf ellipticals are strikingly different. These three sets of models allow
for the blue, low-mass dwarf ellipticals to potentially have a range of
ages from 2 to 14 Gyr, and [Fe/H] from about $-$1 to $-$0.3.
Optical $I$-band SBF measurements of blue dwarf ellipticals 
(Mieske et al. 2006) also show a large spread of fluctuation magnitudes 
and younger implied ages.
Since each set of models was computed with the aim of understanding
something different---the role of $\alpha$-enhancement in early-type galaxies
for the BaSTI models (Lee et al.\ 2010) and the TP-AGB phase for the
SPoT and Padova models (Raimondo 2009; Cantiello et al.\ 2007, 2012;
Lee et al.\ 2010)---the differences shown here provide the starting
point for future detailed comparison of the near-IR properties of 
actual red giant and AGB stars in unresolved stellar populations 
with model predictions. 
We have chosen to show model comparisons using \Mbarj\ plotted against
\gz\ computed using the individual galaxy optical SBF distances. The
general trends and conclusions are similar when the models are
compared to \Mbarh\ or plotted against \JH.

It is beyond the scope of this study to provide a critical analysis of the
strengths and shortcomings of each of these sets of models. For the purposes
of this study, we
conclude that the IR SBF distance calibration is robust when applied to
old, red, metal-rich galaxies like the giant ellipticals typically found in
environments like Virgo and Fornax, and even to some with intermediate-age 
populations and somewhat bluer colors. The bluer dwarf ellipticals, on
the other hand, provide important new observational constraints that should
be of interest to researchers constructing the next generation of stellar
population models. It is clear
that better constraints from data such as these will be valuable
in better defining the properties of young, metal-poor populations, and
the brightness of the TP-AGB stars within these populations.

\section{Recommendations for Measuring IR SBF Distances with WFC3}
WFC3 on the \hst\ makes it possible to measure IR SBF magnitudes at relatively large
distances in modest exposure times.
Based on our experience with the Fornax and Virgo
cluster calibration data presented here, as well tests with WFC3/IR observations in
the Coma cluster from \hst\ program GO-11711 (see Blakeslee 2013) and with the 
instrument exposure time calculator, we provide the following guidelines to help
other astronomers plan WFC3 SBF observations and make use of existing data in the
\hst\ archive.

\begin{enumerate}
\item The fluctuations are brighter at \Hir\ than at \Jir,
but the \Jir\ filter is signficantly wider than \Hir\ (by 0.8 mag), which 
largely cancels out the brightness advantage. 
The image quality is slightly better at \Jir\ than \Hir, and the
background is slightly lower. The net effect is that exposure times to
achieve a particular SBF $S/N$ ratio is about the same in the two filters,
but the ability to detect and remove contaminating point sources (primarily globular
clusters) is better at \Jir, and there is less scatter in the calibration.
We therefore advise choosing \Jir\ over \Hir\ when possible.
The broad F140W filter would be a good alternative, unhampered by the
1.083$\,\mu$m He emission line in the upper atmosphere, 
but that bandpass remains uncalibrated for SBF.

\item Typical one-orbit exposure times (${\sim}2400$\,s) suffice for
  measuring distances in either \Jir\ or \Hir\ out to about 80~Mpc.
This distance limit is imposed by the point source sensitivity 
required to detect and remove globular clusters from the image,
with the goal of reaching within $\sim\,$1~mag of the peak of the globular cluster
luminosity function in \Jir, or $\sim\,$1.5~mag of the peak in \Hir, for which the
fluctuations are relatively brighter (Jensen et al.\ 1998).
The SBF signal itself can be detected to much larger distances
(beyond 100~Mpc) in a single orbit, but the large correction for
contaminating point sources would then dominate the uncertainty.

\item Exposure times for more distant galaxies should be scaled to achieve a
point source sensitivity sufficient to detect and remove globular clusters
1~to 1.5~mag brighter than the peak of the globular cluster luminosity function. 
The exposure time needed to detect the stochastic fluctuations that comprise the SBF
signal goes as $d^2$, but the time required to detect the globular clusters 
increases significantly faster,
scaling as $d^3$ to $d^4$, because of the bright background on
which the globular clusters are superimposed. 

\item To avoid issues with correlated noise, do not use the default \hst\
pipeline-combined images.  Use the \emph{flt} files without correcting 
the WFC3/IR field distortion. 
It may be possible to recreate the \emph{drz} files using
the lanczos3 kernel in the \emph{astrodrizzle} package; this
approach was not tested for WFC3/IR as part of this study, but has been 
used successfully in the past for ACS data (cf. Cantiello et al. 2005; 
Mei et al. 2005a).

\item Because SBF magnitudes depend on the properties of the stellar populations,
high-quality color data are essential to determine accurate distances. 
For the most accurate SBF distances, one should target giant elliptical and S0
galaxies with old stellar populations, for which \gz\ colors are greater than 1.2 or
\JH\ are greater than 0.22 AB mag.
The population variations at bluer colors are too great
for robust distance measurements. If possible, it is best to use ACS \gz\ colors,
but WFC3/IR \JH\ colors are an acceptable alternative.  If necessary, other
color indices may be translated to \gz\ or \JH\ using well-constrained empirical
or model relations for old, metal-rich populations.
\end{enumerate}

\section{Summary}

We have measured \Jir\ and \Hir\ SBF magnitudes and \JH\ colors for 16 early-type
galaxies in the Virgo and Fornax clusters observed with WFC3/IR.  All of these
galaxies have previously measured SBF distances and \gz\ colors from ACS.  
We find that the luminous red galaxies in the sample follow linear relations between
absolute SBF magnitude and optical or IR color;
SBF distances to such galaxies can be determined within a statistical 
uncertainty of 5\%
using the calibration relations that we have presented in 
Equations~(\ref{eq:j_gz}) through~(\ref{eq:h_jh}).
The systematic uncertainty of this calibration is ${\sim}10$\% due to the
uncertainty in the Cepheid calibration on which this work is based.
Stellar population models are not consistent enough to provide a direct calibration
of the IR SBF technique accurate to 10\%, but they are valuable for inferring
age and metallicity trends in and among the sample galaxies.
The scatter in SBF magnitude among bluer galaxies is large, indicating
that a wider variety of stellar populations dominate the light from these 
galaxies and that a simple broadband color does not adequately parameterize the 
complexities inherent in such populations.
Some galaxies, particularly those of intermediate color, clearly have younger 
(4 to 8 Gyr) populations, likely with AGB stars that enhance the IR fluctuation 
amplitude. Bluer galaxies, primarily dwarf ellipticals, may have 
very-low metallicities and/or younger ages, with the interpretations
varying among different sets of stellar population models.
Finally, we have provided practical advice to guide researchers interested
in undertaking SBF measurements with WFC3/IR.

\acknowledgments
Based on observations made with the NASA/ESA \emph{Hubble Space Telescope}, obtained 
at the Space Telescope Science Institute, which is operated by the Association of 
Universities for Research in Astronomy, Inc., under NASA contract NAS 5-26555. 
These observations are associated with program \#11712. 
Additional data from program \#11570 was obtained from the Data Arcive.
J. Jensen, Z. Gibson, and N. Boyer acknowledge the support of the Utah Valley 
University Scholarly Activities Committee. 
Z. Gibson was also supported by the Utah NASA 
Space Grant Consortium award NNX10AJ77H.  
M. Cantiello is grateful for support from the 
FSE-Abruzzo ``Sapere e Crescita'' project and the PRIN-INAF-2014 grant 
(P.I. Gisella Clementini). H. Cho acknowledges support from the 
National Research Foundation of Korea to the Center for Galaxy Evolution 
Research.  
This research has made use of the NASA/IPAC Extragalactic Database (NED) 
which is operated by the Jet Propulsion Laboratory, California Institute of 
Technology, under contract with the National Aeronautics and Space 
Administration. 



\end{document}